%% file: main.tex
\documentclass[longauth]{aa}

\usepackage{graphicx}
\usepackage{txfonts}
\usepackage{lipsum}
\usepackage{subcaption}         
\usepackage{lscape}             
\usepackage{placeins}           
                                
\usepackage{xurl}
\usepackage{xcolor} 
\usepackage[colorlinks=true,allcolors=blue]{hyperref}

\usepackage{booktabs}

\usepackage{orcidlink}

\newcommand{\SVOM}{\textit{SVOM}}
\newcommand{\Swift}{\textit{Swift}}

\newcommand{\EP}{\textit{Einstein Probe}}

\begin{document}

   \title{GRB 241113B: the first SVOM and EP gamma-ray burst}

   \subtitle{A soft GRB likely seen on-axis, from soft X-rays to gamma-rays}


    \author{
        {D.} Adrien\orcidlink{0009-0002-9333-0182}\inst{1}\corrauth{dylan.adrien@cea.fr},
        {D.} Turpin\orcidlink{0000-0003-1835-1522}\inst{1}\email{damien.turpin@cea.fr}, 
        {T.} Maiolino\orcidlink{0000-0002-4918-7182}\inst{2}\email{tais.maiolino@umontpellier.fr},
        {F.} Daigne\orcidlink{0000-0002-4918-7182}\inst{3,4}\email{daigne@iap.fr},
        {F.} Piron\orcidlink{0000-0001-6885-7156}\inst{2}\email{piron@in2p3.fr},
        {Z.-H.} Yao\inst{5}\email{zhyao@bao.ac.cn},
        {Q.} Wu\orcidlink{0009-0002-9275-715X}\inst{5,6}\email{qywu@bao.ac.cn},
        {C.-Y}. Dai\orcidlink{0000-0002-0170-0741}\inst{5,6}\email{cydai@smail.nju.edu.cn},
        {A. V.} Filippenko\orcidlink{0000-0003-3460-0103}\inst{7}\email{afilippenko@berkeley.edu},
        {B.} Schneider\inst{8}\email{benjamin.schneider@lam.fr},
        {B.} Kumar\orcidlink{0000-0001-7225-2475}\inst{9}\email{brajesh@ynu.edu.cn},
        {B.} Cordier\orcidlink{0009-0004-2732-5865}\inst{1}\email{bertrand.cordier@cea.fr},
        {F.} Cangemi\inst{10}\email{cangemi@apc.in2p3.fr},
        {A.} Coleiro\orcidlink{0000-0003-0860-440X}\inst{10}\email{coleiro@apc.in2p3.fr},
        {M. G.} Bernardini\inst{2,11}\email{bernardini@lupm.in2p3.fr},
        {N.} Dagoneau\orcidlink{0000-0002-1361-2562}\inst{12}\email{nicolas.dagoneau@cea.fr},
        {Y.-L.} Qiu\inst{5}\email{qiuyl@nao.cas.cn},
        {L.-P.} Xin\inst{5}\email{xlp@nao.cas.cn},
        {H.-L.} Li\inst{5}\email{lhl@nao.cas.cn},
        {H.} Sun\inst{5,6}\email{hsun@nao.cas.cn},
        {X.-P.} Xu\inst{5,6}\email{xuxp@bao.ac.cn},
        {Z.-Y.} Liu\inst{5,6}\email{ustclzy@mail.ustc.edu.cn},
        {M.-Q.} Huang\inst{5,6}\email{hmengqiu@mail.ustc.edu.cn},
        {M.-H.} Zhang\orcidlink{0009-0006-5706-0364}\inst{5,6}\email{zhangmh@bao.ac.cn},
        {W.} Yuan\inst{5,6}\email{wmy@nao.cas.cn},
        {W.-K.} Zheng\inst{7}\email{weikang@berkeley.edu},
        {G.} Mo\orcidlink{}\inst{13,14}\email{gmo@mit.edu},
        {X.} Liu\orcidlink{0000-0003-1295-2909}\inst{9}\email{x.liu@ynu.edu.cn},
        {Y.} Fang\orcidlink{0009-0006-1010-1325}\inst{9}\email{fangyuan@ynu.edu.cn},
        {G.-W.} Du\orcidlink{0000-0002-8109-7152}\inst{9}\email{dugking@ynu.edu.cn},
        {S.} Guillot\orcidlink{0000-0002-6449-106X}\inst{15}\email{sebastien.guillot@utoulouse.fr},
        {X.-W.} Yu\orcidlink{0000-0002-5881-335X}\inst{16}\email{xywang@nju.edu.cn},
        {R.} Leticia Becerra\orcidlink{0000-0002-0216-3415}\inst{17}\email{rbecerra@astro.unam.mx},
        {J.-Y.} Wei\inst{5,18}\email{wjy@nao.cas.cn},
        {S.} Basa\inst{8}\email{stephane.basa@lam.fr},
        {J.} Wang\inst{5}\email{wj@nao.cas.cn},
        {L.} Bouchet\inst{15}\email{laurent.bouchet@utoulouse.fr},
        {X.-H.} Han\orcidlink{0000-0002-6107-0147}\inst{5}\email{hxh@nao.cas.cn},
        {M.} Brunet\orcidlink{0009-0002-6036-9873}\inst{15}\email{marius.brunet@utoulouse.fr},
        {C.} Wu\orcidlink{0009-0001-7024-3863}\inst{5}\email{cwu@nao.cas.cn},
        {S.} Crepaldi\inst{19}\email{stefano.crepaldi.external@cnes.fr},
        {H.-B.} Cai\inst{5}\email{chb@nao.cas.cn},
        {O.} Godet\orcidlink{0000-0001-7635-9544}\inst{15}\email{olivier.godet@utoulouse.fr},
        {Y.} Xu\inst{5}\email{yxu@nao.cas.cn},
        {D.} G{\"o}tz\inst{1}\email{diego.gotz@cea.fr},
        {P.-P.} Zhang\inst{5}\email{ppzhang@nao.cas.cn},
        {A.} Saccardi\orcidlink{0000-0002-6950-4587}\inst{1}\email{andrea.saccardi@cea.fr},
        {Y.-J.} Xiao\orcidlink{0009-0001-1514-3528}\inst{5}\email{yjxiao@bao.ac.cn},
        {S.} Vergani\orcidlink{0000-0001-9398-4907}\inst{3,20}\email{susanna.vergani@obspm.fr},
        {S.-N.} Zhang\orcidlink{0000-0001-5586-1017}\inst{21}\email{zhangsn@ihep.ac.cn},
        {S.-L.} Xiong\orcidlink{0000-0002-4771-7653}\inst{21}\email{xiongsl@ihep.ac.cn},
        {Y.-W.} Dong\orcidlink{0000-0003-3882-8316}\inst{21}\email{dongyw@ihep.ac.cn},
        {J.-T.} Liu\inst{21}\email{jtliu@ihep.ac.cn},
    }

   \institute{CEA Paris-Saclay, Irfu/D{\'e}partement d’Astrophysique- AIM, 91191 Gif-sur-Yvette, France
       \and 
       Laboratoire Univers et Particules de Montpellier, Universit{\'e} Montpellier, CNRS/IN2P3, F-34095 Montpellier, France
       \and 
       Sorbonne Universit{\'e}, CNRS, UMR 7095, Institut d’Astrophysique de Paris, 98 Bis bd Arago, 75014, Paris, France
       \and 
       Institut Universitaire de France
       \and 
       National Astronomical Observatories, Chinese Academy of Sciences, Beijing 100012
       \and 
       School of Astronomy and Space Science, University of Chinese Academy of Sciences, Beijing 100049, China
       \and 
       Department of Astronomy, University of California, Berkeley, CA 94720-3411, USA
       \and 
       Aix Marseille University, CNRS, CNES, LAM, Marseille, France
       \and 
       South-Western Institute for Astronomy Research, Yunnan University, Kunming 650504, China;  Yunnan Key Laboratory of Survey Science, Kunming 650500, China
       \and 
       Universit{\'e} Paris Cit{\'e}, CNRS, Astroparticule et Cosmologie, F-75013 Paris, France
       \and 
       INAF – Osservatorio Astronomico di Brera, Via Bianchi 46, I23807 Merate, LC, Italy
       \and 
       Universit{\'e} Paris-Saclay, CEA, Département d’{\'E}lectronique des D{\'e}tecteurs et d’Informatique pour la Physique, 91191, Gif-sur-Yvette, France
       \and 
       The Observatories of the Carnegie Institution for Science, 813 Santa Barbara St., Pasadena, CA 91101, USA
       \and 
       Department of Astronomy, California Institute of Technology, 1216 E California Blvd, Pasadena, CA 91125, USA
       \and 
       University of Toulouse, CNES, CNRS, IRAP
       \and 
       School of Astronomy and Space Science, Nanjing University, Nanjing 210023, China
       \and 
       Universidad Nacional Aut{\'o}noma de M\'exico. Instituto de Astronom{\'i}a. A.P. 70-264, 04510. Ciudad de M{\'e}xico, M{\'e}xico.
       \and 
       China Guangxi Key Laboratory for Relativistic Astrophysics, School of Physical Science and Technology, Guangxi University, Nanning 530004, China
       \and 
       Centre National d’{\'E}tudes Spatiales, Centre Spatial de Toulouse,Toulouse Cedex 9, France
       \and 
       LUX, Observatoire de Paris, Universit´e PSL, CNRS, Sorbonne Universit´e, Meudon, 92190, France
       \and 
       Key Laboratory of Particle Astrophysics, Institute of High Energy Physics, Chinese Academy of Sciences, Beijing100049, China
   }

\authorrunning{D. Adrien et al.}

   \date{Received Oct 29, 2026}

 
  \abstract
   {In 2024, two new missions were launched to detect high-energy transients. \EP\ ({\it EP}) and \SVOM\ aim to observe soft transients from X-ray rich gamma-ray bursts (XRR), X-ray flashes (XRF), to fast X-ray transients (FXT).}
   {We studied the case of GRB 241113B, the first soft gamma-ray burst (GRB) jointly detected by \EP\ and \SVOM\ from soft X-rays to gamma-rays (0.5--1000~keV) and evaluated the classification method traditionally applied to soft GRBs.} 
   {We performed time-averaged spectral and temporal analyses of the prompt emission from GRB 241113B over the 0.5–1000 keV energy range. 
   While no redshift was measured, we tentatively constrain its range from the lack of significant dust absorption in the blue afterglow and the deep host-galaxy non-detection in Legacy Survey DR10 archival images. We also fitted a phenomenological and physical model to multiwavelength afterglow data using a smooth broken power law and a forward shock with top-hat jet structure propagating either in  the interstellar medium \textit{(ISM)} or a \textit{Wind} medium.}
  {GRB 241113B has an $E^{\text{obs}}_{\text{peak}} \approx 35$~keV and a redshift (\textit{z}) likely between 0.5 and 3.5, with $T_{90,\gamma} \approx 42$~s (4--30~keV). This burst can be classified as either an XRR or XRF, depending on the fluence ratio used. The afterglow is achromatic and featureless with a normal decay ($\alpha \approx -1.40$). The prompt emission does not satisfy the shock-breakout relation, and thus a geometry comprised of a jet is preferred. However, the afterglow fitting favors a jet viewed on-axis dominated by synchrotron emission due to electrons in a slow cooling regime, interacting with a \textit{wind} medium. The bulk Lorentz factor ($\Gamma$) indicates that the motion of the jet is ultrarelativistic ($\Gamma_{0}>100$), which discards the dirty fireball model. Hence, the jet baryon loading should be compatible with that of classical GRBs (C-GRBs).}
   {The prompt spectral properties of GRB 241113B ($\alpha$, $E^{\mathrm{obs}}_{\mathrm{peak}}$, $E_{\gamma,\mathrm{iso}}$) are comparable to those of XRFs/XRRs and are in the continuity of the C-GRB distributions, setting it apart from the extragalatic FXT population solely detected by {\it EP}. Hence, the \SVOM\ mission, with the benefit brought by the instrument ECLAIRs with its high sensitivity and low energy threshold (at 4~keV), is able to detect the potential $\gamma$-ray counterpart of soft transients also observed by {\it EP} and help to differentiate them from FXTs. The latter show the interest of the joint operation of both missions to better study the landscape of soft transients.}


   \maketitle
   \nolinenumbers

\input{Section/0_Introduction}
\input{Section/1_Detection_and_high_energy_data_reduction}

\input{Section/3_Follow-up_observation}
\input{Section/4_Data_analysis}
\input{Section/5_Afterglow_analysis}
\input{Section/6_Discussion}

\input{Section/7_Conclusion}

\bigskip
\bigskip


The authors thank Xiang-Wang Yu and Sébastien Guillot for their valuable corrections and comments. The Space-based multi-band Variable Objects Monitor (SVOM) is a joint Chinese French mission led by the Chinese National Space Administration (CNSA), the French Space Agency (CNES), and the Chinese Academy of Sciences (CAS). We gratefully acknowledge the unwavering support of NSSC, IAM-CAS, XIOPM, NAOC, IHEP, CNES, CEA, CNRS, University of Leicester, and MPE.

This work also makes use of data obtained with the Einstein Probe, a space mission supported by the Strategic Priority Program on Space Science of the Chinese Academy of Sciences, in collaboration with ESA, MPE and CNES (grant XDA15310000). 

Acknowledgments: The Legacy Surveys consist of three individual and complementary projects: the Dark Energy Camera Legacy Survey (DECaLS; Prop. ID \#2014B-0404; PIs David Schlegel and Arjun Dey), the Beijing-Arizona Sky Survey (BASS; NOAO Prop. ID \#2015A-0801; PIs Zhou Xu and Xiaohui Fan), and the Mayall z-band Legacy Survey (MzLS; Prop. ID \#2016A-0453; PI Arjun Dey). DECaLS, BASS, and MzLS together include data obtained, respectively, at the Blanco telescope, Cerro Tololo Inter-American Observatory, NSF’s NOIRLab; the Bok telescope, Steward Observatory, University of Arizona; and the Mayall telescope, Kitt Peak National Observatory, NOIRLab. Pipeline processing and analyses of the data were supported by NOIRLab and the Lawrence Berkeley National Laboratory (LBNL). The Legacy Surveys project is honored to be permitted to conduct astronomical research on Iolkam Du’ag (Kitt Peak), a mountain with particular significance to the Tohono O’odham Nation.

NOIRLab is operated by the Association of Universities for Research in Astronomy (AURA) under a cooperative agreement with the U.S National Science Foundation (NSF). LBNL is managed by the Regents of the University of California under contract to the U.S. Department of Energy.

This project used data obtained with the Dark Energy Camera (DECam), which was constructed by the Dark Energy Survey (DES) collaboration. Funding for the DES Projects has been provided by the U.S. Department of Energy, the U.S. National Science Foundation, the Ministry of Science and Education of Spain, the Science and Technology Facilities Council of the United Kingdom, the Higher Education Funding Council for England, the National Center for Supercomputing Applications at the University of Illinois at Urbana-Champaign, the Kavli Institute of Cosmological Physics at the University of Chicago, Center for Cosmology and Astro-Particle Physics at the Ohio State University, the Mitchell Institute for Fundamental Physics and Astronomy at Texas A\&M University, Financiadora de Estudos e Projetos, Fundacao Carlos Chagas Filho de Amparo, Financiadora de Estudos e Projetos, Fundacao Carlos Chagas Filho de Amparo a Pesquisa do Estado do Rio de Janeiro, Conselho Nacional de Desenvolvimento Cientifico e Tecnologico and the Ministerio da Ciencia, Tecnologia e Inovacao, the Deutsche Forschungsgemeinschaft and the Collaborating Institutions in the Dark Energy Survey. The Collaborating Institutions are Argonne National Laboratory, the University of California at Santa Cruz, the University of Cambridge, Centro de Investigaciones Energeticas, Medioambientales y Tecnologicas-Madrid, the University of Chicago, University College London, the DES-Brazil Consortium, the University of Edinburgh, the Eidgenossische Technische Hochschule (ETH) Zurich, Fermi National Accelerator Laboratory, the University of Illinois at Urbana-Champaign, the Institut de Ciencies de l’Espai (IEEC/CSIC), the Institut de Fisica d’Altes Energies, Lawrence Berkeley National Laboratory, the Ludwig Maximilians Universitat Munchen and the associated Excellence Cluster Universe, the University of Michigan, NSF’s NOIRLab, the University of Nottingham, the Ohio State University, the University of Pennsylvania, the University of Portsmouth, SLAC National Accelerator Laboratory, Stanford University, the University of Sussex, and Texas A\&M University.

BASS is a key project of the Telescope Access Program (TAP), which has been funded by the National Astronomical Observatories of China, the Chinese Academy of Sciences (the Strategic Priority Research Program ``The Emergence of Cosmological Structures,'' grant \# XDB09000000), and the Special Fund for Astronomy from the Ministry of Finance. The BASS is also supported by the External Cooperation Program of Chinese Academy of Sciences (grant \# 114A11KYSB20160057), and the Chinese National Natural Science Foundation (grant \# 12120101003, \# 11433005).

The Legacy Survey team makes use of data products from the {\it Near-Earth Object Wide-field Infrared Survey Explorer (NEOWISE)}, which is a project of the Jet Propulsion Laboratory/California Institute of Technology. {\it NEOWISE} is funded by the National Aeronautics and Space Administration (NASA).

The Legacy Surveys imaging of the DESI footprint is supported by the Director, Office of Science, Office of High Energy Physics of the U.S. Department of Energy under Contract No. DE-AC02-05CH1123, by the National Energy Research Scientific Computing Center, a DOE Office of Science User Facility under the same contract; and by the U.S. National Science Foundation, Division of Astronomical Sciences, under Contract No. AST-0950945 to NOAO.

KAIT and its ongoing operation were made possible by donations from Sun Microsystems, Inc., the Hewlett-Packard Company, AutoScope Corporation, Lick Observatory, the U.S. NSF,      
the University of California, the Sylvia \& Jim Katzman Foundation, and the TABASGO Foundation. Research at Lick Observatory is partially supported by a generous gift from Google.

A.V.F.’s research group at U.C. Berkeley acknowledges financial assistance from the Christopher R. Redlich Fund, Gary and Cynthia Bengier, Clark and Sharon Winslow, Alan Eustace and Kathy Kwan (W.Z. is a Bengier-Winslow-Eustace Specialist in Astronomy), and numerous other donors.

Mephisto is developed at and operated by the South-Western Institute for Astronomy Research of Yunnan University (SWIFAR-YNU), funded by the ``Yunnan University Development Plan for World-Class University" and ``Yunnan University Development Plan for World-Class Astronomy Discipline". The Mephisto team acknowledges support from the Key Laboratory of Survey Science of Yunnan Province (202449CE340002), the ``Science \& Technology Champion Project'' (202005AB160002), and from two ``Team Projects" -- the ``Top Team'' (202305AT350002) and the ``Innovation Team'' (202105AE160021), funded by the ``Yunnan Revitalization Talent Support Program". B.K. acknowledges the support from the ``Special Project for High-End Foreign Experts", Xingdian Funding from Yunnan Province, and two National Key Research and Development Programs of China (2024YFA1611603, 2024YFA1611604).

WINTER’s construction is made possible by the National Science Foundation under MRI grant number AST-1828470 with early operations supported by AST-2206730. Significant support for WINTER also comes from the California Institute of Technology, the Caltech Optical Observatories, the Bruno Rossi Fund of the MIT Kavli Institute for Astrophysics and Space Research, the David and Lucille Packard Foundation, and the MIT Department of Physics and School of Science.


%

\bibliographystyle{bibtex/aa.bst} 
\bibliography{biblio.bib} %

\input{Section/8_Appendix}

\end{document}

%% file: Section/0_Introduction.tex
\section{Introduction}
Gamma-ray bursts (GRBs) consist of gamma-ray prompt emission that lasts from a few milliseconds up to thousands of seconds with an extragalactic origin \citep{Metzger1997}. The GRB prompt emission arises mainly from internal shocks within a high relativistic jet \citep{Rees1992,Sari_1997,Piran1999,Kumar_Zhang2015}. This jet can then interact with the circumburst medium (CSM) and produce external shocks from which accelerated electrons will cool by radiating broadband X-ray to radio synchrotron emission called the afterglow \citep{Meszaros_1997, Sari_1998,Zhang_2006}.

GRBs are known to originate from at least two types of central-engine stellar explosions: (1) mergers of compact objects such as two neutron stars (NS-NS) or a neutron star and a black hole (NS-BH) produce Type I GRBs \citep{Zhang2018,Gehrels_2005,Abbott_2017}, often characterized by short spikes ($T<2$~s) of hard gamma-ray emission, and (2) massive-star core collapse preferentially associated with Type Ic-BL supernovae (SNe) induce Type II GRBs \citep{Galama_1998,Campana_2006,Li2025}, which are usually spectrally softer and with much longer gamma-ray prompt emission ($T\geq2$~s) than the Type I bursts \citep{Kouveliotou1993}.

In this paper we mainly focus on Type II GRBs, which exhibit a wide distribution of $E^{\rm obs}_{\rm peak}$; the peak of the observed spectral energy distribution (SED). Because the $E^{\rm obs}_{\rm peak}$ ganges from a few to thousands of keV, the softest bursts are harder to detect with gamma-ray instruments which typically explore the sky above 10~--~20~keV \citep{Lien_2016}. However, thanks to the soft band detection capabilities of the HETE-2 mission \citep[FREGATE: (6 -- 400)~keV and WXM: (2--25)~keV; ][]{Ricker_2003}, \cite{Sakamoto2005} proposed an empirical and crude spectral subclassification of Type II GRBs (that we will call hereafter the \textit{Sakamoto subclassification}), to discriminate soft from hard bursts, by computing the X-ray to gamma-ray fluence ratio (SR = S$_{X}$/S$_{\gamma}$). \cite{Sakamoto2005} separated long GRBs into three subclasses, namely classical-GRB (C-GRB), X-ray rich GRB (XRR), and X-ray flash (XRF). This subclassification is defined as follows:

\begin{align*} 
 & S_{X}(2\text{--}30\text{ keV})/S_{\gamma}(30\text{--}400\text{ keV}) \leq 0.32 && \text{(C-GRB)}, \\
 0.32 < {} & S_{X}(2\text{--}30\text{ keV})/S_{\gamma}(30\text{--}400\text{ keV}) \leq 1.0 && \text{(XRR)}, \tag{\theequation} \stepcounter{equation} \\
 & S_{X}(2\text{--}30\text{ keV})/S_{\gamma}(30\text{--}400\text{ keV}) > 1.0 && \text{(XRF)} 
\end{align*}
\label{eq1}

With this subclassification, \cite{Sakamoto2005} suggested that XRFs are bursts with $E^{\rm obs}_{\rm peak} \leq 30\ \text{keV}$, and XRRs have $E^{\rm obs}_{\rm peak}$ of 30--100 keV. Moreover, it was found that soft bursts have a smaller peak flux ($F_{p}$), and thus are fainter than C-GRBs. These results were supported by \cite{Sakamoto2008}, where in addition to the low luminosity of soft bursts, it was shown that XRF/XRR X-ray afterglows also differ from C-GRB X-ray afterglows. \cite{Sakamoto2008} also converted Eq. (\ref{eq1}) to the narrower bandpass of the \Swift/BAT (15--150 keV; \citealt{Swift/BAT}):
\begin{align*} 
 & S_{X}(25\text{--}50\text{ keV})/S_{\gamma}(50\text{--}100\text{ keV}) \leq 0.72 && \text{(C-GRB)}, \\
 0.72 < {} & S_{X}(25\text{--}50\text{ keV})/S_{\gamma}(50\text{--}100\text{ keV}) \leq 1.32 && \text{(XRR)}, \tag{\theequation} \stepcounter{equation} \\
 & S_{X}(25\text{--}50\text{ keV})/S_{\gamma}(50\text{--}100\text{ keV}) > 1.32 && \text{(XRF)} 
\end{align*}
\label{eq2}
For this second iteration of the subclassification, the boundary between XRFs and XRRs ($S_{X}(25\text{--}50\text{ keV})/S_{\gamma}(50\text{--}100\text{ keV}) =1.32$) is defined as the fluence ratio obtained with a GRB spectrum that follows a Band function \citep{Band1993} with $\alpha = -1$, $\beta = -2.5$, and $E^{\rm obs}_{\rm peak} = 30$ keV, whereas the limit between XRRs and C-GRBs is the fluence ratio of a burst spectrum, with $\alpha = -1$, $\beta = -2.5$, and $E^{\rm obs}_{\rm peak} = 100$ keV. 

The prompt emission duration ($T_{90}$) and redshift ($z$) distribution of the XRFs/XRRs is somewhat similar to the C-GRBs, but differ mainly in prompt energetics ($E_{\gamma, {\rm iso}}$), rate, and luminosity \citep{Bi_2018,Katsukura_2020}. 
XRRs/XRFs have less energetic prompt emission than most C-GRBs (XRF, $\log(E_{\gamma,{\rm iso}})\approx 51$; XRR, $\log(E_{\gamma,{\rm iso}})= 52.6^{+0.6}_{-0.9}$; C-GRB, $\log(E_{\gamma,{\rm iso}})= 53.1^{+0.7}_{-1.0}$; \citealp{Xu_2025}). In addition, they tend to cluster at the tail of the Type II GRBs distribution in $E_{\text{peak},z}$--$E_{\gamma,\text{iso}}$ (known as the \textit{Amati relation}; see Figure 10 of \citealt{Bi_2018}).
XRRs/XRFs are less luminous than hard Type II bursts \citep[see Figure 9(b) of][]{Xu_2025}, and some XRFs could be related to low-luminosity GRBs (LL-GRBs; $L<10^{50}$ erg s$^{-1}$; \citealt{Virgili_2009}). According to \cite{Virgili_2009}, LL-GRBs, and some XRFs, could form a distinct population from C-GRBs, and this second population would be mandatory to fit the long GRB luminosity function at low luminosity. However, more recent studies have shown that a triple-power-law model could better fit the luminosity function of long GRBs without the extra need of LL-GRBs being a distinct population \citep{Sun_2015}.

A computation of the rate (number of events in the universe per year) of each Type II GRB subclasses has been made by \cite{Katsukura_2020}, giving $f_{\rm C-GRB}$ = 570 $\pm$ 36, $f_{\rm XRR}$ = 3031 $\pm$ 53, and $f_{\rm XRF}$ = 968 $\pm$ 45 events yr$^{-1}$. Using the cosmic GRB rate function found by \cite{Lien_2014} and the standard cosmological parameters given by \cite{Planck_Collaboration_2020}, one can translate the aforementioned rate to a local event rate density, which gives $\rho_{0,{\rm C-GRB}}$ = $0.39\pm0.03$ Gpc$^{-3}$ yr$^{-1}$, $\rho_{0,{\rm XRR}}$ = $0.66\pm0.03$ Gpc$^{-3}$ yr$^{-1}$, and $\rho_{0,{\rm XRF}}$ = $2.07\pm0.04$ Gpc$^{-3}$ yr$^{-1}$.

Because Type II GRB subclasses usually follow classical correlations (such as the Amati, Yonetoku, and $L_{p}$--$T_{90,{\rm rest}}$--$E_{\rm iso}$ relations), they likely form a continuous distribution where XRRs and XRFs represent the low-energy tail \citep{Xu_2025}. Therefore, several authors \citep{Sakamoto2008,Katsukura_2020} argue that XRRs and XRFs are not novel events requiring unique progenitors, but rather the same phenomenon as C-GRBs operating under different physical conditions. This is demonstrated by the variety of scenarios proposed to explain them, including the following:

\begin{enumerate}[\textbf{\arabic*})]
    \item \textbf{Cosmological effects:} 
    \begin{itemize}
        \item \textit{High Redshift:} \cite{Heise2001} proposed that XRFs could be high redshift GRBs ($z > 5$). But more recent observations have shown that soft GRBs have the same redshift distribution than C-GRBs, with 90\% of XRRs and XRFs having $z < 2.8$ \citep{Bi_2018}.
    \end{itemize}
    \item \textbf{Geometrical effects: }
    \begin{itemize}
        \item \textit{Varying jet opening angle:} \citet{Lamb_2005} showed that a top-hat jet model (constant emissivity per unit solid angle within the jet) with varying opening angle could explain the distribution and continuity from C-GRBs to XRFs (at least for the \textit{HETE}-2 sample with measured redshifts). Nevertheless, this model predicts that XRFs should have wider jets under the strong assumption (challenged by more recent studies; \citealp{Wang_2018}) that the corrected beamed energy $E_{\gamma}$ is similar among all GRBs --- i.e., a standard reservoir of energy \citep{Bloom_2003}.
        
        \item \textit{Varying observing angle:} In the case of a uniform jet (the variation of energy per solid angle is constant), \cite{Yamazaki2002} proposed that XRFs should be bursts that have a viewing angle ($\theta_{\nu}$) within $\Delta\theta \lesssim \theta_{\nu}\lesssim 2 \Delta\theta$, while XRRs/C-GRBs have a $\theta_{\nu}$ within $0\lesssim\theta_{\nu}\lesssim\Delta\theta$, where $\Delta\theta$ is the radiative jet half-opening angle \citep{Yamazaki2004}. Other types such as a universal jet and a quasiuniversal jet viewed off-axis are also believed to be capable of spawning XRFs/XRRs \citep{Zhang2004,Rossi2002}.     
    \end{itemize}
    \item \textbf{Jet content: }
    \begin{itemize}
        \item \textit{Heavy baryon load:} GRB jets that are heavily loaded with baryons should have a bulk Lorentz factor ($\Gamma$) lower than 100 and create a ``dirty fireball'' which would seed fewer $\gamma$-ray photons than the typical GRB, resulting in a higher observed fluence ratio SR $= S_{X}/S_{\gamma}$ \citep{Huang2002}. 
        \item \textit{Small contrast of shell's Lorentz factor:} Contrary to the ``dirty fireball,'' the ``clean fireball'' model shows that XRFs/XRRs can be produced from jets with $\Gamma>>300$, but with a small contrast of Lorentz factors between the colliding shells in order to reduce the efficiency of internal shocks to generate fewer $\gamma$-ray photons \citep{Mochkovitch_2004}.
    \end{itemize}
    \item \textbf{Dissipation process:}
    \begin{itemize}
        \item \textit{Photosperic model:} Some XRFs exhibit thermal components in their prompt emission (XRF 060218, \citealt{Campana_2006}; XRF 241001A, \citealt{Schneider2026}). Based on that observation, \citet{peers_2006} suggested that such a thermal component should arise through kinetic dissipation happening close to the thermal photosphere.
    \end{itemize}
\end{enumerate}

To date soft GRBs, including XRRs and XRFs, remain sparsely explored, as previous missions typically had energy thresholds of 10–30 keV, limiting their capability to detect these class of events. Nevertheless, these events constitute a critical missing component for completing our understanding of central-engine–driven explosion mechanisms and the diversity of associated astrophysical scenarios. Since 2024, two new missions, \textit{Einstein Probe} ({\it EP}) \citep{Yuan2022} and \textit{SVOM} \citep{Cordier2026}, equipped with X-ray (\textit{EP}/WXT: 0.5--4 keV) and soft $\gamma$-ray (\textit{SVOM}/ECLAIRs: 4--120 keV) instruments optimized for the detection of XRRs and XRFs, have provided enhanced capabilities to investigate these populations. 

In this paper, we report the first joint detection of a soft GRB by \SVOM\ and EP, namely GRB 241113B/EP 241113b. This event highlights the strong synergy between these two missions in probing soft GRBs. In Sections \ref{sec:prompt} and \ref{sec:follow-up}, we present our observations of the prompt and afterglow emission of GRB 241113B. The results of the prompt emission spectral analysis are presented in Section \ref{sec:prompt_analysis}, where we favour a Type II GRB classification and classify the event as an XRR/XRF according to the Sakamoto subclassification.  Section \ref{sec:afterglow_analysis}  performs afterglow modeling to derive additional physical properties of GRB 241113B and compares them with those of the Type II classical GRB (C-GRB) population. In Section \ref{sec:discussion}, we discuss how GRB 241113B may challenge the Sakamoto subclassification, use the relation from \citet{Nakar_2012} to test the shock-breakout (SBO) scenario, compare this burst to other off-axis GRBs, and examine its properties in the context of recent XRR/XRF detections by \SVOM\ and {\it EP}. Our conclusions are presented in Section \ref{sec:conclusion}. Throughout the paper, we adopt a flat $\Lambda CDM$ cosmology of the \citet{Planck_Collaboration_2020} with $\Omega_{m}=0.315$, $\Omega_{\Lambda}=0.685$, and H$_{0}=67.4$ km s$^{-1}$ Mpc$^{-1}$. All uncertainties are quoted at the 1~$\sigma$ level or 68\% confidence interval (C.I.) and all upper limits are reported at 3~$\sigma$ level, unless stated otherwise.

%% file: Section/1_Detection_and_high_energy_data_reduction.tex
\begin{figure*}[t]
\centering
\includegraphics[width=1.0\textwidth]{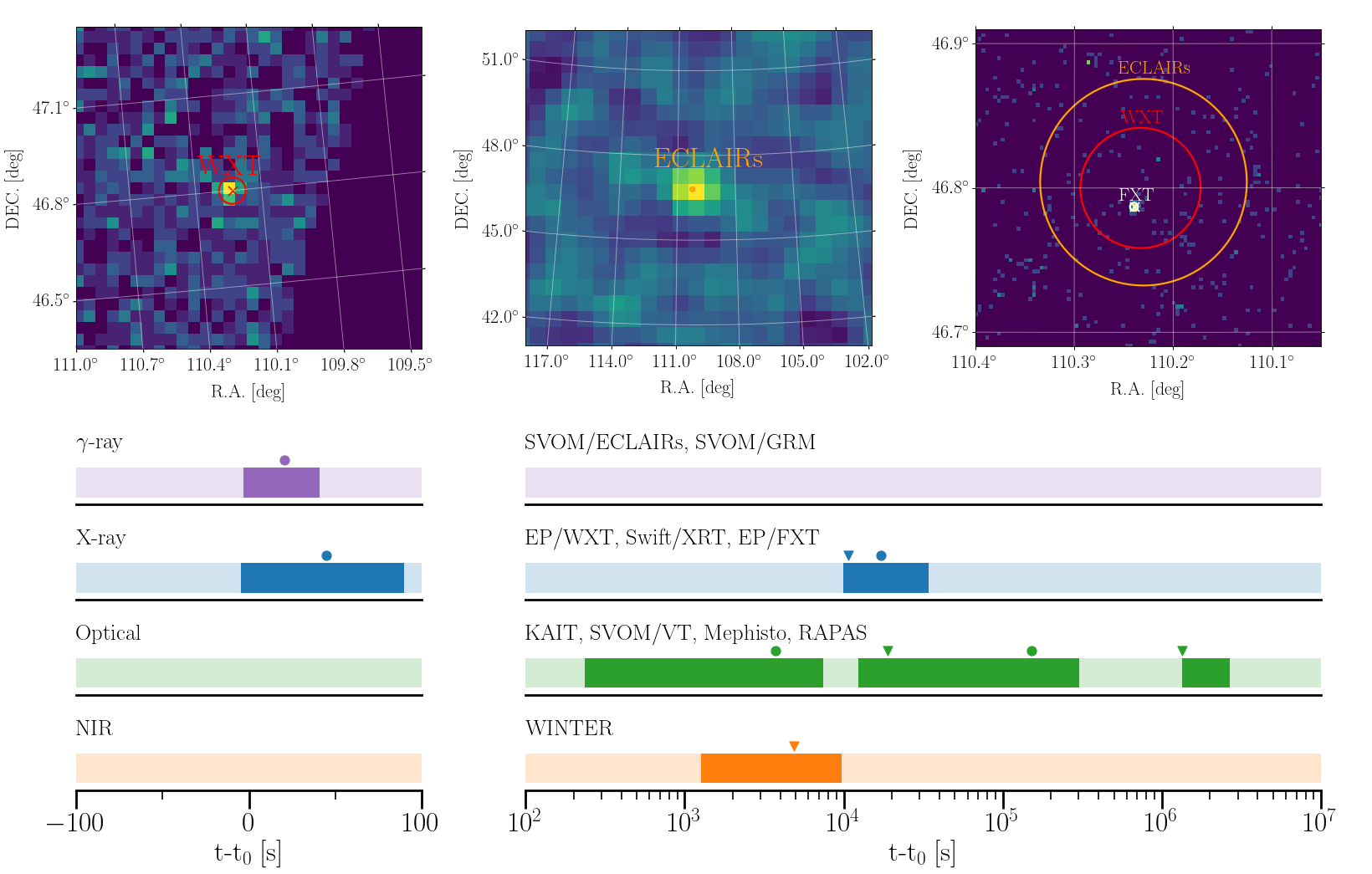}
    \caption{Timeline of the multiwavelength observations of GRB 241113B/EP 241113b. Each wavelength is represented by an axis and a color. Dots indicate actual detections, while triangle are upper limits. Titles above each axis correspond to the wavelength considered and the list of telescopes operating in that wavelength that provided data. The top panels represent the positions of the transient in {\it EP}/WXT, \SVOM/ECLAIRs, and {\it EP}/FXT (left to right). The red, orange, and white circles correspond to localization error boxes for GRB 241113B, as determined by {\it EP}/WXT, {\it SVOM}/ECLAIRs, and {\it EP}/FXT, respectively.}
    \label{fig: Timeline}
\end{figure*}

\section{Prompt emission}

\label{sec:prompt}
\subsection{High-energy detection}

GRB 241113B triggered {\it EP}/Wide field X-ray telescope (WXT) (in \textit{CMOS23}) at 2024-11-13T11:22:50 (UTC dates and times are used throughout this paper;  \citealt{EPGCNGRB241113B}), a few arc seconds from its border. \SVOM/ECLAIRs then triggered 15 seconds later at 2024-11-13T11:23:05  \citep{ECLAIRsGCNGRB241113B} and the burst is located $\sim$25 deg from its optical axis. It was also detected by the \SVOM\ gamma-ray burst monitor (GRM) \citep{GCN_GRM_GRB241113B} and positioned close to the second GRM detector (GRD2) field of view (FoV) center. The significance of the detection in each instrument computed for different energy bands (see Appendix \ref{appendix: significance}) shows that GRB 241113B is only marginally detected above 50 keV, and the signal is faint in GRD3, thus we decided to exclude the latter in our analysis. The lack of a significant detection of GRB 241113B in energy bands above 50 keV is indicative of the softness of this burst. The high energy detection timeline and the burst localization in {\it EP}/WXT and \SVOM/ECLAIRs FoV are shown on Figure \ref{fig: Timeline}.

\subsubsection{\EP/WXT}

{\it EP}/WXT data were processed using a data-reduction pipeline and the calibration database (CALDB) specifically designed for WXT (Liu et al., in prep.). The CALDB incorporates results both from on-ground and in-orbit calibration observations \citep{Cheng2024}. Event energies were calibrated on the pulse-invariant (PI) channel scale using the gain and bias values from the CALDB. Bad and flaring pixels were flagged, and event grades were assigned. Events with grades 0--12 and no anomalous flags were selected to construct the cleaned event file. From the cleaned event file, the source and background (light curve and spectrum) were extracted using \texttt{XSELECT V2.5b}\footnote{\url{https://heasarc.gsfc.nasa.gov/docs/software/ftools/xselect/}}, from a circle with a radius of $9'$ and an annulus of $18'$ and $36'$, respectively. The events selected to build the light curve are the one between Pulse Height Amplitude (PHA) channels 50--500, which translates to 0.5--4~keV in energy.

\subsubsection{\SVOM/ECLAIRs and \SVOM/GRM}

The ECLGRM X-band pipeline \citep{Piron2026} version 2.6.0 was used to compute the background-subtracted count light curves (BSCLCs) and durations of the GRB observed by both ECLAIRs and GRM instruments. The same pipeline was also employed to extract the source spectrum from ECLAIRs, as well as from the GRD 1 and 2 detectors of the GRM instrument. For ECLAIRs, the GRB duration was estimated by considering only events originating from pixels illuminated by the GRB in the detector plane, thereby reducing background effects. We found that the $T_{90}$ for ECLAIRs is best constrained when computed in the 4--30 keV energy range, in which the burst is significantly detected.

The BSCLCs were created in the default time interval ($t_0- 108.8$~s to $t_0+403.2$~s), and binned with a 1.0~s interval during the GRB window to capture detailed temporal variations, and with a 5.0~s interval outside this window, helping to reduce noise when less detailed sampling is needed. The best light curve background modeling was given by the so-called ModelE, a polynomial function which takes into account the angle between the detector’s main axis and the satellite-Earth direction. The time-integrated spectra were extracted between $t_0-8.22$~s and $t_0+35.67$~s.

\subsection{Light Curves}

\begin{figure*}[t]
\centering
\includegraphics[width=1.0\textwidth]{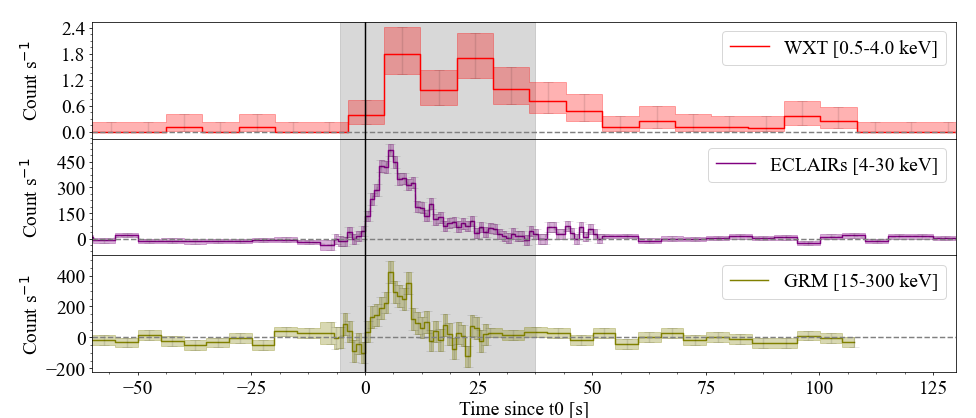}
    \caption{Background subtracted count-rate light curves of WXT, ECLAIRs, and GRM. The WXT light curve is in 8~s bins, while ECLAIRs and GRM are binned to 5~s  oustide the GRB window and 1~s inside. The GRB windows are defined, independently, for both ECLAIRs and GRM, using a Bayesian blocks algorithm inside the ECLGRM X-band pipeline. The gray shadow is the interval chosen to perform time-averaged spectral analysis, which corresponds to the ECLAIRs $T_{90}$ interval $\pm$ 5~s, starting from 2024-11-13T11:23:01.941 to 2024-11-13T11:23:45.832.}
    \label{fig: prompt_lc}
\end{figure*}

The light curves of this burst (see Figure \ref{fig: prompt_lc}) exhibit a single asymmetrical peak, composed of a fast rise and exponential decay (FRED pulse). All signals are well aligned to the ECLAIRs trigger time (2024-11-13T11:23:07.600), which we took as the reference time ($t_{0}$) for the rest of the analysis. In Figure \ref{fig: prompt_lc}, the {\it EP}/WXT light curve was rebinned to 8.0~s bins owing to the low net count rate ($\sim 0.1$ count~s$^{-1}$). The gray region in Figure \ref{fig: prompt_lc} corresponds to the time interval chosen to perform the spectral extraction in each instrument in order to build the joint time-integrated spectrum (spectral analysis and results are given in Section \ref{sec:prompt_analysis}). This time interval is derived from the ECLAIRs $T_{90}$ interval $\pm$ 5~s ([2024-11-13T11:23:01.941 -- 2024-11-13T11:23:45.832]; see Section \ref{section: T90}).

%% file: Section/3_Follow-up_observation.tex
\section{Follow-up observations}
\label{sec:follow-up}
Follow-up observations of GRB~241113B spanned from 103~s to 15~days post-\SVOM/ECLAIRs trigger, covering the X-ray, near-infrared (NIR), and optical bands. A timeline summarizing these observations, or the lack thereof, is presented in Figure~\ref{fig: Timeline}. All afterglow data points used in this work are recorded in Table~\ref{tab:followup_obs}, and all the data reductions leading to these measurements are presented in subsequent subsections.

\subsection{X-ray observations}

\subsubsection{\textit{Swift}/XRT}

On 2024-11-13T14:21:22.374 ($\sim 3.0$ hr after $t_{0}$), we performed a first target-of-opportunity (ToO) observation with the \Swift/XRT instrument with 1.5 ks exposure \citep{Evans2024}. No uncataloged X-ray source was detected either in the \SVOM/ECLAIRs or the \EP/WXT error boxes. Assuming a typical GRB spectrum, this places a quite constraining 0.3--10 keV flux upper limit of $\mathrm{2.4\times10^{-13}~erg~cm^{-2}~s^{-1}}$ \citep{Page2024}.

\subsubsection{\textit{Einstein Probe}/FXT}

On 2024-11-13 15:43:11 ($\sim$4.3 hr after $t_{0}$), a second ToO observation was performed using the \EP\ Follow-up X-ray telescope (FXT) instrument \citep[\textit{EP}/FXT;][]{chen2025a} with 3~ks exposure. An uncataloged X-ray counterpart was identified within the \SVOM/ECLAIRs and \EP/WXT error box at R.A. (J2000) = 110.2397$^\circ$, Dec. = +46.7870$^\circ$; see Figure \ref{fig: Timeline}. The source was detected at moderate signal-to-noise ratio (SNR) in the FXT A (SNR $\approx 6.1$) and FXT B (SNR $\approx 9.1$) telescopes and was later confirmed as being the X-ray afterglow of GRB 241113B thanks to the identification of its fast fading behavior at optical wavelengths \citep{Turpin2024}. The spectral results of this observation are given in section \ref{sec:xaglow_spec_analysis}.


\subsection{Optical observations}

\subsubsection{KAIT}
The 0.76\,m Katzman Automatic Imaging Telescope \citep[KAIT;][]{Filippenko2001}, located at Lick Observatory (near San Jose, California, USA), automatically responded to GRB~241113B starting at 11:24:48, 103~s after ECLAIRs trigger. Observations were conducted in a $3 \times 3$ tiling mode using a {\it Clear} filter \citep[approximately equivalent to $R$; see][]{LiWeidong2003} and lasting for about two hours. In one of the KAIT tiling fields, we discovered the optical afterglow \citep{Zheng2024GCN38232} at R.A. (J2000) = 110.2405$^\circ$, Dec. = +46.7867$^\circ$ ($\sigma = 0.5''$) which was later confirmed by the \SVOM/VT \citep{VTTeam2024GCN38306}.

All KAIT images were reduced using a custom pipeline\footnote{\url{https://github.com/benstahl92/LOSSPhotPypeline}} detailed by \citet[][]{Stahl2019}. Point-spread-function (PSF) photometry was obtained using {\tt DAOPHOT} \citep[][]{Stetson1987} from the {\tt IDL} Astronomy Users Library\footnote{\url{http://idlastro.gsfc.nasa.gov/}}. Photometric calibration was achieved using local standard stars from the Pan-STARRS1 Surveys \citep{Schlafly_2012}, with magnitudes transformed into the Landolt standard system \citep{Landolt_1992} using the relations from \cite{Tonry_2012}. All {\it Clear} filter images were calibrated to the $R$ band. Detailed photometric results are summarized in Table~\ref{tab:followup_obs}. 

\subsubsection{\SVOM/VT}
The Visible Telescope (VT;~\citealt{Qiu2026}) is the 44~cm optical instrument onboard  \SVOM, providing simultaneous two-band imaging in VT$\_B$ (400--650 nm) and VT$\_R$ (650--1000 nm) for rapid monitoring of GRBs and other transients. \SVOM/VT initiated ToO observations of GRB~241113B starting at 2024-11-13T14:23:24, about 3.0~hr after the trigger. In total, VT executed five consecutive orbits of follow-up observations.

The VT images were processed following standard optical imaging reduction procedures using IRAF v2.16~\citep{Tody1986, Tody1993}. Individual frames were corrected for instrumental signatures, including bias/dark subtraction as applicable and flat-fielding, and then astrometrically registered to a common world coordinate system~\citep{Li2026}. For each band, the reduced exposures were aligned and combined to produce deep stacked images, on which source detection and photometry were performed at the afterglow position. Aperture photometry was done using a circular aperture of radius $1.14''$ (1.5~pixels), with the local sky background estimated from an annulus with inner/outer radii of 7.6--11.4$''$ (10--15~pixels); the measured fluxes were then aperture-corrected to a $7.6''$ (10-pixel) radius using growth curves derived from bright, isolated field stars. The photometric zero-points \citep{Yao2026} were determined using calibrated field stars in the same frames, and the resulting magnitudes were reported on the AB system \citep{Oke_1983}. The afterglow was detected in the VT$\_R$ band in the first three orbits, while only limiting magnitudes were obtained in the last two orbits; in the VT$\_B$ band, the source was detected in the first two orbits and was not detected thereafter, for which we report limiting magnitudes in the remaining three orbits (see Table~\ref{tab:followup_obs}).

\subsubsection{Mephisto}

We performed simultaneous $vrz$-band observations of GRB~241113B with the Multi-channel Photometric Survey Telescope (Mephisto) located at Lijiang Observatory \citep{Yuan2020}. The facility is operated by the South-Western Institute for Astronomy Research, Yunnan University. Mephisto is equipped with three-channel CCD cameras: blue ($uv$), yellow ($gr$), and red ($iz$) channels. It can do imaging simultaneously in $ugi$ or $vrz$ optical bands. The wavelength coverages in the $u$, $v$, $g$, $r$, $i$, and $z$ bands are 320--365, 365--405, 480--580, 580--680, 775--900, and 900--1050 nm, respectively \citep[see][for the Mephisto filter system]{ChenX-2024C, Yuan-PeiY}. The Mephisto observations started at 16:17:01 (on 2024 November 13) in the $vrz$ bands, and six frames of 300~s each were obtained.

Data processing was carried out with a dedicated pipeline developed for Mephisto, which includes the preprocessing steps (bias and dark subtraction, flat-fielding corrections, and cosmic-ray removal). The photometric calibration was done on the stacked images by the synthetic photometry technique and utilizing the Gaia BP/RP spectra \citep{Huang-2024ApJS, Xiao-2024ApJS}. The calibration accuracy is better than $\sim 0.03$ mag in $u$, $\sim 0.01$ mag in $v$, and $\lesssim 0.005$ mag in $griz$ \citep[the detailed methodology is presented by][]{ChenX-2024C, Zhou_2026}. No credible detection was found at the optical transient location, and our limiting magnitudes are listed in Table~\ref{tab:followup_obs}. 

\subsubsection{WINTER}
We observed GRB 241113B in the near-infrared (NIR) $J$ band with the Wide-field Infrared Transient explorer \citep[WINTER;][]{2020SPIE11447E..9KL, 2024SPIE13096E..3JF}, a camera mounted on the 1~m telescope at Palomar Observatory. Data were reduced using the standard WINTER pipeline built using \texttt{mirar} \citep{mirar}. The observations began at 2024-11-13T12:14:05 (51 min after the GRB) and consisted of $30 \times 120$~s exposures. In our stacked images, we did not detect any NIR transient candidate at the position reported by KAIT and VT down to $J = 19.3$ mag (3$\sigma$ confidence level).

\subsubsection{Search for Host Galaxy in Archival Survey Catalog}
\label{sec:search_for_host}

\begin{figure*}[t]
    \centering
    \includegraphics[trim=0 20 0 0,width=1.0\textwidth]{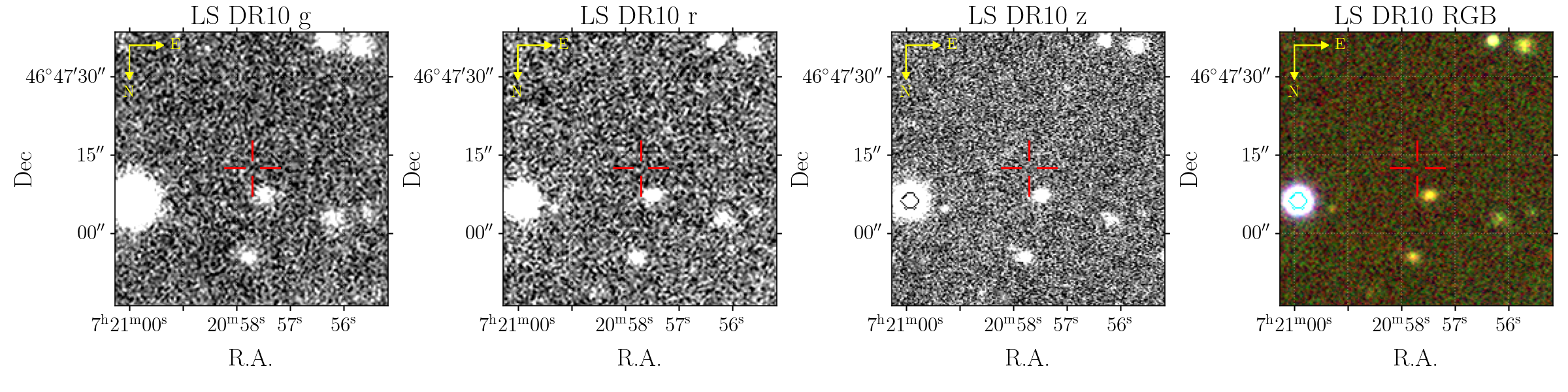}
    \caption{Cutouts from the Legacy Survey DR10 in the $g$, $r$, and $z$ bands (and an RGB composite image) centered at the optical position given by KAIT at R.A., Dec. (J2000) = 110.2405$^\circ$, +46.7867$^\circ$. No host galaxy could be identify down to $g = 24.5$ mag, $r = 23.9$ mag, and $z = 23.2$ mag.}
    \label{fig:LS10_host_search}
\end{figure*}

From our follow-up campaign, we were unable to identify any contribution from a potential host galaxy in the afterglow light curve. Host galaxies can be used to measure the redshift of the GRB through the identification of emission lines in their optical spectra. Therefore, we searched for a potential faint host galaxy in the Legacy Survey DR10 \citep[LS10; ][]{Dey2019} image catalog. We selected $g$-, $r$-, and $z$-band cutouts centered at the optical position reported by KAIT; see Figure \ref{fig:LS10_host_search}. No source was detected at the position of GRB 241113B, and forced photometry was performed. We conclude that there is no host galaxy down to $g = 24.5$ mag, $r = 23.9$ mag, and $z = 23.2$ mag at the 5$\sigma$ confidence level.

\subsection{Redshift Constraints}
\label{sec:redshift_constraints}
No redshift could be obtained for GRB 241113B; however, we can constrain its possible range thanks to our optical observations and archival data.\\
\subsubsection{Upper limit estimation} First, the optical afterglow was detected by both VT channels (B: 450--650 nm, R: 650--1000 nm) with a moderate red color index $VT_{B} - VT_{R} \approx 0.2$ mag, suggesting negligible absorption at bluer wavelengths and very likely that the $VT_{B}$ flux is not strongly suppressed by the Lyman-alpha (Ly$_\alpha$) break. A simple projection of the Ly$_\alpha$ break in the middle of the $VT_B$ wavelength coverage (which would have severely reddened the color index by strongly suppressing the flux at bluer wavelengths) gives us a conservative upper limit for the redshift of GRB 241113B, $z~\lesssim~3.5$. 
\subsubsection{Lower limit estimation}
The second constraint is given by the deep flux limit of the potential host galaxy of GRB 241113B we have extracted from the Legacy Survey DR10; see Section \ref{sec:search_for_host}. We can compare our magnitude upper limit  with the known GRB host-galaxy magnitude distribution as a function of redshift \citep[the TOUGH sample]{Hjorth2012}. Based on this statistical distribution, a GRB host galaxy with $r > 23.9$ mag is unlikely to be located at $z\lesssim 0.5$, as this would require the host to be both unusually low in mass and intrinsically faint. Such a low-redshift, sub-luminous GRB host has previously been identified in the case of GRB 030329, at $z\simeq0.17$ and an approximate apparent magnitude V $\sim$ 22.7 \citep{Fruchter2003,Thone2007}. However, the host galaxy of GRB 241113B would be even more extreme: given the magnitude upper limits derived here, it would be approximately at least a factor of two less luminous than the already sub-luminous host of GRB 030329 if it was localized at the same redshift.\\
Based on all these considerations, we chose to favor a redshift range $0.5 \lesssim z \lesssim 3.5$ for GRB 241113B.

%% file: Section/4_Data_analysis.tex
\section{Prompt emission analysis}
\label{sec:prompt_analysis}
\subsection{Prompt duration ($T_{90}$)}
\label{section: T90}

The ECLGRM X-band pipeline compute the $T_{90}$ duration, for ECLAIRs and GRM, by performing a Poisson resampling of the observed count light curve (CLC) and generating 300 simulated background-subtracted cumulative count light curves (BSCCLCs). For each BSCCLC, the pipeline computes its plateau, as well as the $t_{95}$ and $t_{05}$ times, when 95\% and 5\% of the GRB's total counts are reached, respectively. $T_{90}$ is then estimated as the median of the resulting $t_{95}$--$t_{05}$ distribution.

For WXT, a Bayesian-block analysis was first applied to the background-subtracted light curve, in the 0.5--4 keV energy range, to define the WXT emission interval, from the first to the last block with SNR $\geq 3$ \citep{Scargle2013}. Within this interval, $T_{90}$ was calculated as the time interval containing 90\% of the total net counts, from 5\% to 95\% of the cumulative counts. Its uncertainty was estimated by Poisson-resampling the source and background light curves and recomputing $T_{90}$ for each realization. The quoted uncertainty corresponds to the central 68\% interval of the resulting $T_{90}$ distribution, as reported in Table \ref{tab:t90_comparison}. 

\begin{table}[h!]
    \centering
    
    \renewcommand{\arraystretch}{1.2} 

    \setlength{\tabcolsep}{2pt} 

    \footnotesize
    \begin{tabular}{lcccc}
        \toprule
        Instrument & Energy Band & $T_{90}$  & $T_{05} - t_{0}$ & $T_{95} - t_{0}$ \\
                   & (keV)       & (s) & (s) & (s)     \\
        \midrule
        WXT        & 0.5--4      & $89.5_{-21.0}^{+2.5}$ & -4.60 & 94.1\\
        \addlinespace
        ECLAIRs    & 4--30       & $42.1^{+4.2}_{-4.0}$  & -3.37 &   44.2 \\
        ECLAIRs    & 4--120       & $39.5^{+30.0}_{-5.7}$  & -3.67 & 67.7       \\
        \addlinespace
        GRM        & 15--550     & $10.0^{+27.5}_{-1.5}$ & -3.19 & 35.0      \\
        GRM        & 50--300     & $5.3^{+10.3}_{-2.2}$ & -3.77 & 14.0      \\
        \bottomrule
    \end{tabular}
    \caption{$T_{90}$ durations of GRB 241113B measured for each instruments. Note that the energy bands listed correspond to the specific ranges used for analysis, whereas the full operating ranges are 0.5--4 keV (WXT), 4--120 keV (ECLAIRs), and 15--5000 keV (GRM). Here $T_{05}$ - $t_{0}$  ($T_{95}$ - $t_{0}$) represents the position of the $T_{05}$ ($T_{95}$) lowest (highest) bound with respect to $t_{0}$.}
    \label{tab:t90_comparison}
\end{table}

We compared $T_{90}$ in all instruments and found that it is significantly higher in WXT compared to ECLAIRs and GRM. This behavior is expected and is related to the fact that GRB pulses are narrower at higher energy \citep{Fenimore_1995}, which we investigate in Section \ref{subsec: T90-Energy}. But, one can also notice that $T_{90}$ in the same instruments can differ from one energy band to another, which gives unconstrained $T_{90}$; for example, in ECLAIRs we have $T_{90,4-30~{\rm keV}} = 42.1^{+4.2}_{-4.0}~s$  and $T_{90,4-120~{\rm keV}} = 39.5^{+30.0}_{-5.7}$. This change of $T_{90}$ with respect to energy band in ECLAIRs, but also GRM, is not only due to the peak narrowing with energy as in \citet{Fenimore_1995}, but also to the softness and faintness of GRB 241113B, for which too few photons are detected above 30~keV to properly estimate $T_{90}$ for higher energies. In the case of ECLAIRs and GRM the best energy band to estimate the burst duration, in order to have the most constrained value of $T_{90}$, is 4--30~keV. Hence, ECLAIRs being the most sensitive in this energy range, we define ECLAIRs $T_{90}$ at 4--30~keV ($T_{90,4-30~keV} = 42.1^{+4.2}_{-4.0}$) as the reference duration of GRB 241113B prompt ($\gamma$-ray) emission for the rest of the analysis.

We computed the boundary of the $T_{90}$ interval with respect to $t_{0}$ and report it in Table \ref{tab:t90_comparison}. The {\it EP}/WXT $T_{05}$ starts $\sim 1$~s before the ECLAIRs one, and thus there are no signs of a precursor in the {\it EP}/WXT data, as shown by the light curve (see Figure \ref{fig: prompt_lc}). On the other hand, the $T_{95}$ in WXT is about 90~s away from $t_{0}$, higher than the difference ($T_{95} - t_{0}$) found in ECLAIRs ($\sim 44$~s) and GRM ($\sim 35$~s). This delay in $T_{95}$ means that thanks to its sensitivity, \textit{EP/WXT} was able to detect the soft extended emission of GRB 241113B that is missed by less sensitive $\gamma$-ray instruments. Furthermore, the light curve (see Figure \ref{fig: prompt_lc}) shows that the soft extended emission detected by WXT is the continuation of the exponential decay of the main pulse. Therefore, the signal detected by all the instruments, from soft X-ray (WXT) to $\gamma$-ray (ECLAIRs and GRM), exhibits self similarity over 4 decades of energy \citep{Nemiroff_1994}, only affected by a combination of instrumental and observational selection effects. These effects  make a burst appear longer or shorter depending on the instrument sensitivity/bandpass and the spectral shape that is being sampled, as discussed by \cite{Virgili_2012}, and narrower at high energies \citep{Fenimore_1995}.

\subsection{$T_{90}$--Energy relation}
\label{subsec: T90-Energy}

\begin{figure}
    \centering
    \includegraphics[width=1.0\columnwidth]{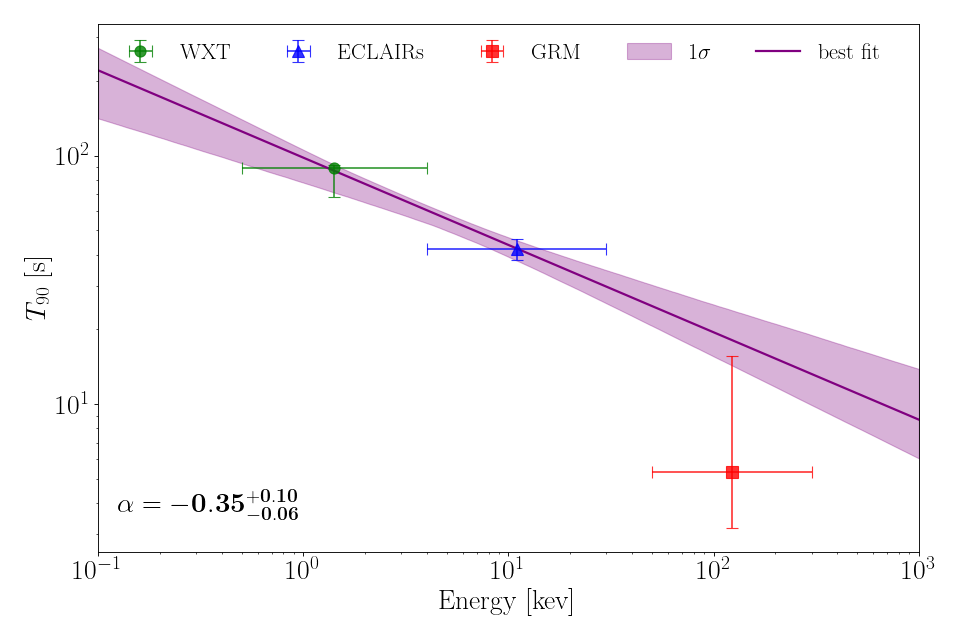}
    \caption{$T_{90}$--energy relation fitted using measurements reported in Table \ref{tab:t90_comparison}. WXT, ECLAIRs, and GRM points corresponds to the $T_{90}$ obtained in the 0.5--4, 4--30, and 50--300 keV energy bands, respectively. Only the errors over $T_{90}$ are used for the fit, and are assumed to be Gaussian. Here, $\alpha$ corresponds to the power-law exponent use to fit the data.}
    \label{fig:T90vsEnergy}
\end{figure}

We investigated the dependence of the GRB duration ($T_{90}$) with respect to the energy band by fitting for each instrument a $T_{90}$ estimate that comes from the nonoverlapping energy band:  $T_{90,~0.5\text{--}4~{\rm keV}}$, $T_{90,~4\text{--}30~{\rm keV}}$ and $T_{90,~50\text{--}300~{\rm keV}}$, for WXT , ECLAIRs, and GRM, respectively (see Table  \ref{tab:t90_comparison}), and used maximum-likelihood estimation (MLE) to fit the  model 
\begin{equation}
    \log{T_{90}} = \alpha\log\left(\frac{E}{50~{\rm keV}}\right) +\log{K_{50}}.
\end{equation}
We found $\alpha = -0.35^{+0.10}_{-0.06}$ and $K_{50}=54.2^{+2.8}_{-4.2}$ (see Figure \ref{fig:T90vsEnergy}). The value of $\alpha$ is consistent with the results reported by \cite{Norris_1995} and \cite{Fenimore_1995}, during the BATSE era, who observed that $\alpha \approx 0.41$ and $\sim$0.37--0.46, respectively. The duration vs. energy relation was also revisited by \cite{Bissaldi_2011}, using {\it Fermi}/GBM bursts, and found that this relation extends up to MeV energies with a power-law exponent that is $\sim$0.3--0.4.
\subsection{Spectral analysis}

\label{sec:prompt_spec_analysis}

\begin{table*}[t]
\centering
\caption{Joint fit}

\renewcommand{\arraystretch}{1.2} 

\setlength{\tabcolsep}{2pt} 

\footnotesize

\resizebox{\textwidth}{!}{%
\begin{tabular}{lccccccccccc}
\toprule
Model & $N^{\rm host}_{\rm H,X}$ $^{(a)}$ & $\alpha$ & $\beta$ & $E^{\rm obs}_{\rm peak}$ & $F_{[4-120]\ {\rm keV}}$ & $F_{[15-150]\ {\rm keV}}$ & $F_{p[0.5-4]\ {\rm keV}}$ $^{(b)}$ & $F_{p[4-120]\ {\rm keV}}$ $^{(c)}$ & STAT/dof $^{(d)}$ & BIC \\ 
 & ($10^{22}$ cm$^{-2}$) & & & (keV) & \multicolumn{1}{c}{(${10^{-8}}$ erg cm$^{-2}$ s$^{-1}$)} & \multicolumn{1}{c}{(${10^{-8}}$ erg cm$^{-2}$ s$^{-1}$)} & \multicolumn{1}{c}{(${10^{-8}}$ erg cm$^{-2}$ s$^{-1}$)} & \multicolumn{1}{c}{(${10^{-8}}$ erg cm$^{-2}$ s$^{-1}$)} & &\\ 
\midrule
PL & $0.75^{+0.16}_{-0.14}$ & $-1.93^{+0.06}_{+0.06}$ & -- & -- & $2.26^{+0.13}_{-0.13}$ &  $1.61^{+0.14}_{-0.13}$ & $5.7^{+4.1}_{-2.7}$ & $6.62^{+0.60}_{-0.60}$ & 304/294 & 326.6 \\
CPL & $0.63^{+0.17}_{-0.16}$ & $-1.80^{+0.16}_{-0.10}$ & -- & $37^{+34}_{-14}$ & $2.22^{+0.14}_{=0.14}$ & $1.54^{+0.15}_{-0.16}$ & $4.8^{+3.3}_{-2.2}$ & $6.50^{+0.61}_{-0.61}$ & 298/293 & 326.8 \\
Band & $0.33^{+0.10}_{-0.06}$ & $-1.14^{+0.16}_{-0.14}$ & $-2.17^{+0.14}_{-0.21}$ & $14.0^{+3.5}_{-2.8}$ & $2.15^{+0.14}_{-0.14}$ & $1.43^{+0.16}_{-0.16}$ & $2.5^{+1.8}_{-1.2}$ & $6.32^{+0.60}_{-0.61}$ & 295/292 & 329.5 \\ 
\bottomrule
\label{tab: joint_spectral_results}
\end{tabular}%
}
\parbox{\textwidth}{
    \vspace{1ex} 
    \footnotesize 
    $^{(a)}$\ Review{Intrinsic absorption (\texttt{"TBabs"}) free to vary within the 90\% confidence interval determined from the {\it EP}/WXT spectrum best fit.} \\
    $^{(b)}$1~s peak flux in the range 0.5--4~keV. \\
    $^{(c)}$1~s peak flux in the range 4--120~keV. \\
    $^{(d)}$STAT: CSTAT + PGSTAT. \\
}
\label{tab: 2}
\end{table*}

ECLAIRs and GRM time-averaged spectra were extracted, using both instruments in counting mode, with the ECLGRM X-band pipeline \citep{Piron2026} version 2.6.0 and were background subtracted using ModelE, on a time interval between $t_0-8.22$~s and $t_0+35.67$~s (roughly $T_{90,ECL}\pm5$~s). The WXT time-averaged spectrum was extracted with \texttt{XSELECT V2.5b} over the same time interval as the ECLAIRs and GRM spectra, and was rebinned using the ftgrouppha\footnote{\url{https://heasarc.gsfc.nasa.gov/docs/software/lheasoft/help/ftgrouppha.html}} method (part of \texttt{HEASoft}); since the WXT spectrum has a low number of counts (50  counts), this avoids having too many background bins with 0 counts which can bias CSTAT \citep{Cash1979}.

The spectral analysis was performed using \texttt{Xspec v12.14.0h} \citep{Arnaud1996} through the Python interface \texttt{PyXspec v2.1.3}. We fitted the GRB 241113B spectrum with three models: a power law (PL), a cutoff power law (CPL), and the \citet{Band1993} function (Band), described by:

\begin{align} 
\text{(PL)} &\quad A(E) = K E^{\alpha}\\[1.0ex]
\text{(CPL)} &\quad A(E) = K E^{\alpha} \exp\left(-\frac{E}{E_{c}}\right)\\[1.0ex]
\text{(Band)} &\quad A(E) = K 
\begin{cases} 
\left(\frac{E}{100}\right)^{\alpha} \exp\left(-\frac{E}{E_c}\right) & \text{if } E < (\alpha - \beta) E_c \\ \\[1.5ex]
\left[\frac{(\alpha - \beta)E_c}{100}\right]^{\alpha - \beta} e^{\beta - \alpha} \left(\frac{E}{100}\right)^{\beta} & \text{if } E \ge (\alpha - \beta) E_c\, 
\end{cases}
\end{align}
\noindent

We also added to the aforementioned models two multiplicative components to estimate the absorption due to ISM. The first one aims at reproducing the Galactic ISM absorption, while the second is used to account for potential additional absorption due to the intergalactic medium (IGM) and intrinsic host-galaxy contributions. Those two components are both labeled, in \texttt{Xspec}, as \texttt{"TBabs"}\footnote{\url{https://heasarc.gsfc.nasa.gov/docs/software/xspec/manual/node318.html}}, which stands for Tuebingen-Boulder ISM absorption model. Finally, we adopted the abundances from \citep{Wilms2000}.

To account for inter-calibration between the missions, we added a constant multiplicative factor (\texttt{"factor"}) to the fitted model. This factor was fixed to 1 for both ECLAIRs and GRM. For \textit{EP}/WXT, we applied a Gaussian prior with a mean of 0.93 and a scatter of 0.11. This prior was derived from joint \textit{EP} and \SVOM\ observations of Cyg X-1, taken one week before GRB 241113B. Measuring the extrapolated 3--6~keV band yielded inter-calibration values of $0.99 \pm 0.08$ and $0.87 \pm 0.10$ for CMOS 14 and CMOS 37, respectively.

Hence, the final fitted model can be written as follow: \texttt{"contstant*TBabs*TBabs*Source"}, where \texttt{"Source"} stands for either PL (\texttt{pow}), CPL (\texttt{cutoffpl}), or Band (\texttt{grbm}).

\subsubsection{Method}

\begin{enumerate}
    \item First, we determine the intrinsic hydrogen column density ($N^{\rm host}_{\rm H,X}$) by fitting the model \texttt{"TBabs * TBabs * pow"} to the WXT time-averaged spectrum. Using the $N_{\rm H}$ procedure of \texttt{HEASoft}\footnote{\url{https://heasarc.gsfc.nasa.gov/docs/software/lheasoft/help/nh.html}}, we found that the Galactic equivalent hydrogen column density is $N^{\rm gal}_{\rm H,X} = 7.20\times{10^{20}}~$cm$^{-2}$ at the GRB position according to the HI4PI survey \citep{HI4PI}. We then found that the intrinsic $N_{\rm H,X}$ is $0.74^{+0.62}_{-0.49}\times{10^{22}}$ cm$^{-2}$ (90\% C.I.; see Appendix \ref{appendix: WXT}).
    For the joint fit we used PGSTAT for ECLAIRs and GRM, and CSTAT for WXT.
    \item We also determined for each model the peak flux by computing the count to flux factor that we then applied to the brightest 1~s bin in the light curve. 
    \item We selected the best model using BIC \citep{BIC1978} and the following related strength of evidence found in \citet{Kass1995} to compare the model in terms of complexity given the available data. According to \citet{Kass1995}, a $2 <$ $\Delta$BIC ($6<$ $\Delta$BIC) means moderate (strong) evidence toward the model with the lower BIC.
    

\end{enumerate}

\subsubsection{Results}
\label{sec: Result of spectral analysis}
\begin{figure}[h]
     \centering
     \includegraphics[width=1.0\columnwidth]{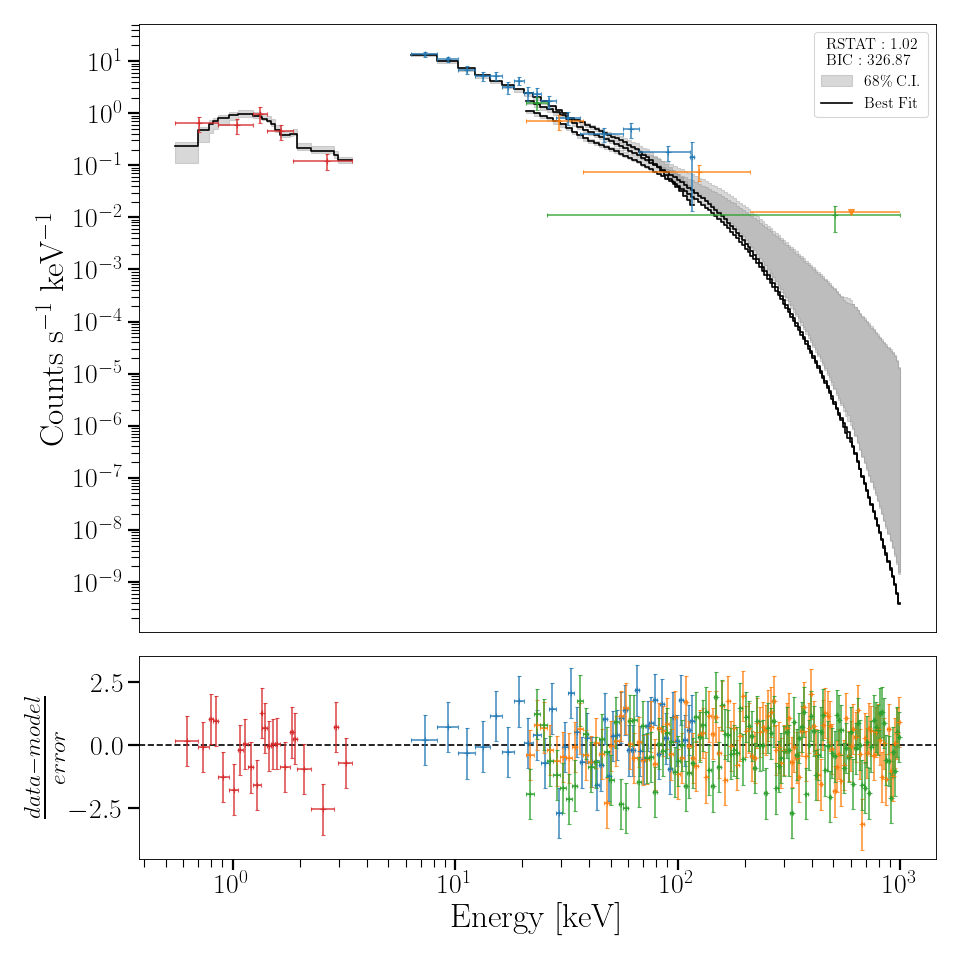}
     \caption[]{\textit{top panel}: EP/WXT (red), ECLAIRs (blue), GRD1 (orange) and GRD2 (green) count rate spectrum. Each spectrum is rebinned to achieve a 3~$\sigma$ signal in each spectral bin for readability, as the true data follow poisson statitstic and bin above 50~keV are mostly background dominated. Triangle correspond to upper limit in the spectral bin, i.e $C_{net} = C_{source}+C_{background}<\sqrt{C_{source}+C_{background}}$. The dark line correspond to the best fit model (CPL), and the grey shadow to the 68\% C.I. (1~$\sigma$) envelop, both convolved by the instruments response and rebinned according to the instrument spectral resolution. \textit{bottom panel}: Residual with respect to the best fit model (CPL).}
     \label{fig:count_rate_spectrum}
\end{figure}

The results of the joint fit of the GRB 241113B time-averaged spectrum are reported in Table \ref{tab: 2}. Evaluating the models using the BIC, we have $\Delta$BIC = BIC$_{\rm Band}$ - BIC$_{\rm PL}$ = 2.9. Therefore, there is positive evidence against the Band function, as its improvement to the fit over the PL model is not significant enough with regard to the number of parameters it introduces. However, for the PL and CPL models we have $\Delta$BIC = BIC$_{\rm CPL}$ - BIC$_{\rm PL}$ = 0.2, and thus both models seem to be indistinguishable. We tried to measure the break significance by computing the likelihood ratio between PL and CPL and found a p-value of 0.014 (2.2 $\sigma$). Hence, we can say that with a confidence beyond the 0.05 threshold: show a marginal preference for a spectral break according to the data, leading us to chose CPL as our primary model. From the CPL model, we estimated GRB 241113B $E^{\rm obs}_{\rm peak}$, to be $37.1^{+34.4}_{-13.7}$ keV (see Figures \ref{fig:count_rate_spectrum} and \ref{fig:theo_spec}).


We computed the fluence ratio defined by \citet{Sakamoto2005} and \citet{Sakamoto2008}, noted respectively as $SR_{1}$ and $SR_{2}$ hereafter, for the different spectral model and report the result in Table \ref{tab: fluence ratio}. We find that $\text{SR}_{1} = S(2-30~{\rm keV})/S(30-400~{\rm keV}) = 1.17^{+0.45}_{-0.28}$ and $\text{SR}_{2} = S(25-50~{\rm keV})/S(50-100~{\rm keV}) = 1.07^{+0.23}_{-0.17}$ with CPL, which would classify this burst as an XRF by \cite{Sakamoto2005} and an XRR by \cite{Sakamoto2008}. We observed that the subclassification of GRB 241113B does not change much according to the model in the case of SR$_{2}$; the burst remains a XRR, except in the case of the Band function where it can marginally be also a XRF. SR$_{1}$ is more sensitive to a change of spectral model; the burst is classified as an XRR with the power-law model, but as an XRF in the case of the CPL and Band models (for more details see Section \ref{sec: sakamoto discussion}). 

\begin{figure*}[htbp]
    \centering
    
    \begin{subfigure}[t]{0.48\textwidth}
        \centering
        \includegraphics[width=\linewidth]{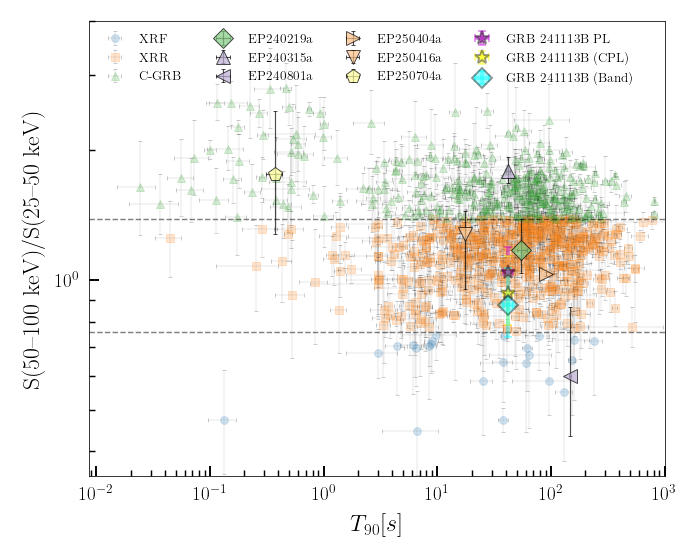}
        \caption{The SR--$T_{90}$ diagram, where the fluence ratio used is the one of \citet{Sakamoto2008}. The location of GRB 241113B is shown in the case of PL, CPL and Band. The green, orange, and blue points slightly transparent in the background are the C-GRBs, XRRs, and XRFs from the BAT3 catalog, respectively \citep{Lien_2016}. The dashed lines represent the separation between XRFs/XRRs and XRRs/C-GRBs as defined by \citet{Sakamoto2008}. EP240219a \citep{Yin_triggering}, EP240315a \citep{Liu_2025}, EP240801a \citep{Jiang_2025}, EP250404a \citep{Yin2}, EP250416a \citep{Zhao_2026} and EP250704a \citep{Li_2026} are also shown for comparison.}
        \label{fig:sakamoto}
    \end{subfigure}%
    \hfill 
    \begin{subfigure}[t]{0.48\textwidth}
        \centering
        \includegraphics[width=\linewidth]{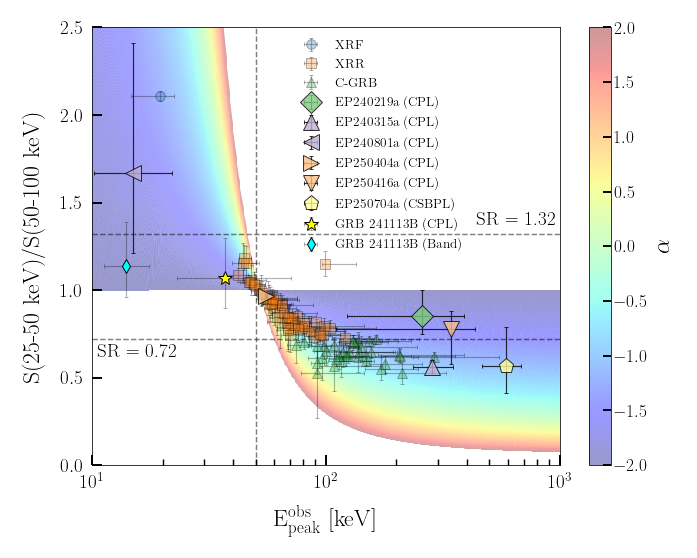}
        \caption{The SR--$E^{\rm obs}_{\rm peak}$ diagram, where the fluence ratio used is the one of \cite{Sakamoto2008}. GRB 241113B is the yellow star, while the green, orange, and yellow points are the C-GRBs, XRRs, and XRFs of the BAT3 catalog \citep{Lien_2016}, respectively. The color map in the background corresponds to parametric CPLs that have $E^{\rm obs}_{\rm peak}\in[1,1000~\rm keV]$ and $\alpha\in[-2,2]$. The dashed horizontal line corresponds to the separation between the different subclasses of long bursts as defined by \citet{Sakamoto2008}, while the vertical one correspond to $E^{\rm obs}_{\rm peak} = 50~\rm keV$. EP240219a \citep{Yin_triggering}, EP240315a \citep{Liu_2025}, EP240801a \citep{Jiang_2025}, EP250404a \citep{Yin2}, EP250416a \citep{Zhao_2026} and EP250704a \citep{Li_2026} are also shown for comparison.}
        \label{fig:sr_vs_epeak}
    \end{subfigure}
    
    \caption{GRB 241113B compared with other XRFs/XRRs/C-GRBs detected by \textit{Swift} as a function of their fluence ratio.}
    \label{fig:combined_figures}
\end{figure*}

We then place GRB 241113B in a hardness vs. $T_{90}$ plane (see Figure \ref{fig:combined_figures}). This object is right in the middle of the distribution of GRBs and is typical of this transition population between C-GRBs and XRFs known as XRRs. With its rather low $E^{\rm obs}_{\rm peak}\sim37~{\rm keV}$ and according to Figure \ref{fig:sr_vs_epeak}, GRB~241113B appears softer than most BAT GRBs best fitted by the CPL model \citep{Lien_2016}.

\begin{table}[h]

\centering
\caption{Fluence ratio}
\renewcommand{\arraystretch}{1.2} 

\resizebox{0.95\columnwidth}{!}{%
\begin{tabular}{l|cc|cc}
\toprule
Model & ${\text{SR}}_{1}$$^{a}$ & subclass$^{a}$ & ${\text{SR}}_{2}$$^{b}$ & subclass$^{b}$  \\ 
\midrule
PL & $0.88^{+0.20}_{-0.16}$ & (XRR/XRF) & $0.96^{+0.14}_{-0.13}$ & (XRR) \\
CPL & $1.17^{+0.45}_{-0.28}$ & (XRF/XRR) & $1.07^{+0.23}_{-0.17}$ & (XRR) \\
Band & $2.62^{+0.51}_{-0.39}$ &  (XRF) & $1.14^{+0.25}_{-0.18}$ & (XRR/XRF) \\ 
\bottomrule
\end{tabular}%
} 
\parbox{0.90\columnwidth}{
    \vspace{1ex} 
    \footnotesize 
    $^a$Fluence ratio and subclassification as defined by \citet{Sakamoto2005}. \\
    $^b$Fluence ratio and subclassification as defined by \citet{Sakamoto2008}. \\
    
}
\label{tab: fluence ratio}
\end{table}

\begin{figure}[htbp]
     \centering
     \includegraphics[width=0.48\textwidth]{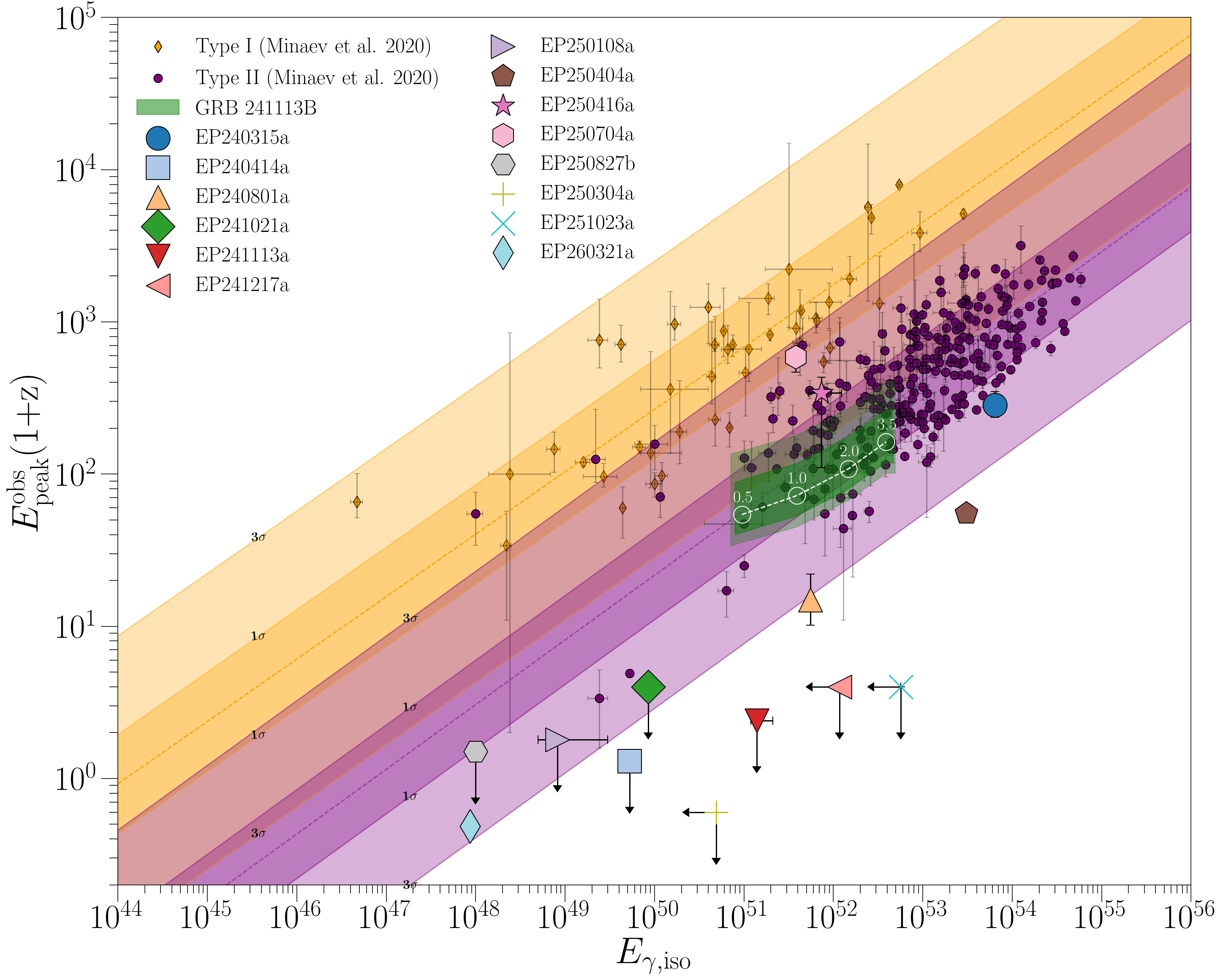}
     \caption{GRB 241113B track in the $E^{\rm src}_{{\rm peak}}$--$E_{\gamma.\rm iso}$ plane (in green) for the estimated range of redshift z $\in [0.5-3.5]$. The orange and purple shadows represent  $1\sigma$ and $3\sigma$ of the Amati correlation for  Type I and Type II GRBs, respectively, given by \citet{Minaev2020}. In addition, others EP transients are show for comparison (EP240219a: \citealt{Yin_triggering}; EP240315a: \citealt{Liu_2025}; EP240414a: \citealt{Sun_2025}; EP240801a: \citealt{Jiang_2025}; EP241021a: \citealt{Shu_2025}; EP241113a: \citealt{Dai2026}; EP241217a: \citealt{Zhou_Hao_2026}; EP250108a: \citealt{Li_2025}; EP250404a: \citealt{Yin2}; EP250416a: \citealt{Zhao_2026}; EP250704a: \citealt{Li_2026}; EP250827b: \citealt{Srinivasaragavan_2026}; EP250304a: \citealt{Cotter_2026}; EP251023a: \citealt{Jiang_2026}; EP260321a: \citealp{Yuan_2026}).}
     \label{fig:amati_relation}
     \label{fig:amati}
\end{figure}

From our joint spectral analysis and the redshift constraints described in Sections \ref{sec:prompt_spec_analysis} and \ref{sec:search_for_host}, respectively, we are able to insert the possible positions of GRB 241113B in the $E^{\rm src}_{{\rm peak}}$--$E_{\rm \gamma,iso}$ plane; see Figure \ref{fig:amati_relation}. This places GRB 241113B in the Type II GRB population. Depending on its redshift, this burst could lie at the faint end of the Type II GRB energy distribution or in the middle of it. We also use the "Energy–Hardness–Duration" (EHD) classification, proposed by \citet{Minaev2020}, that aims at blindly separating Type I from Type II bursts, even if a GRB has no redshift measurements. The EHD classification is defined as:
\begin{equation}
    {\rm EH} = \frac{(E^{\rm src}_{{\rm peak}}/100~\text{keV})}{(E_{\gamma,{\rm iso}}/10^{51}~\text{erg})^{0.4}}\, ,
\end{equation}

\begin{equation}
{\rm EHD}=\frac{\rm EH}{(T^{\rm src}_{90}/1~{\rm s})^{0.5}}=\frac{(E^{\rm src}_{{\rm peak}}/100~{\rm keV})}{(E_{\gamma,{\rm iso}}/10^{51}~\text{erg})^{0.4}(T^{\rm src}_{90}/1~{\rm s})^{0.5}}\, ,
\end{equation}
\noindent
This classification also shows that GRB 241113B is located in the Type II region for all redshifts from 0.5 to 3.5 (see Appendix \ref{appendix: EDH classification}).

%% file: Section/5_Afterglow_analysis.tex
\section{Afterglow emission analysis}
\label{sec:afterglow_analysis}
\subsection{X-ray spectral analysis}
\label{sec:xaglow_spec_analysis}
The \textit{EP}/FXT (A and B) spectra were processed following the standard data-reduction procedure described in the FXT Data Analysis Software Package\footnote{\url{http://epfxt.ihep.ac.cn/analysis}} (FXTDAS v1.10). Given the low SNR of the FXT detection, both count-rate spectra were grouped to a minimum of 1 count per bin using the \textit{grppha} ftools routine. The spectra were then fitted with a simple model \texttt{"TBabs~* TBabs * pow"}. 
As described in Section \ref{sec:prompt_spec_analysis}, the first \texttt{TBabs} parameter is fixed to the Galactic equivalent hydrogen column density $N^{\rm gal}_{H,X} = 7.20 \times 10^{20}~{\rm cm}^{-2}$ while the second \texttt{'TBabs'} component is left free, within the constraints provided from the \textit{EP}/WXT spectral analysis (see Appendix \ref{appendix: WXT}). We used the CSTAT likelihood to fit our data and obtain a best fit with a photon index $\Gamma = 4.2^{+1.5}_{-1.2}$ and $N^{\rm host}_{H,X} = 0.51^{+0.39}_{-0.20}\times~10^{22}~{\rm cm}^{-2}$. This leads to an absorbed (unabsorbed) 0.3--10 keV flux $F_X = 0.7^{+0.3}_{-0.2}~(10.3^{+64.1}_{-4.2}) \times 10^{-13}$~erg~cm$^{-2}$~s$^{-1}$.


\subsection{Phenomenology of the GRB 241113B Afterglow}
\label{sec:pheno_fit}
\begin{figure*}
    \centering
    \includegraphics[width=1.0\textwidth]{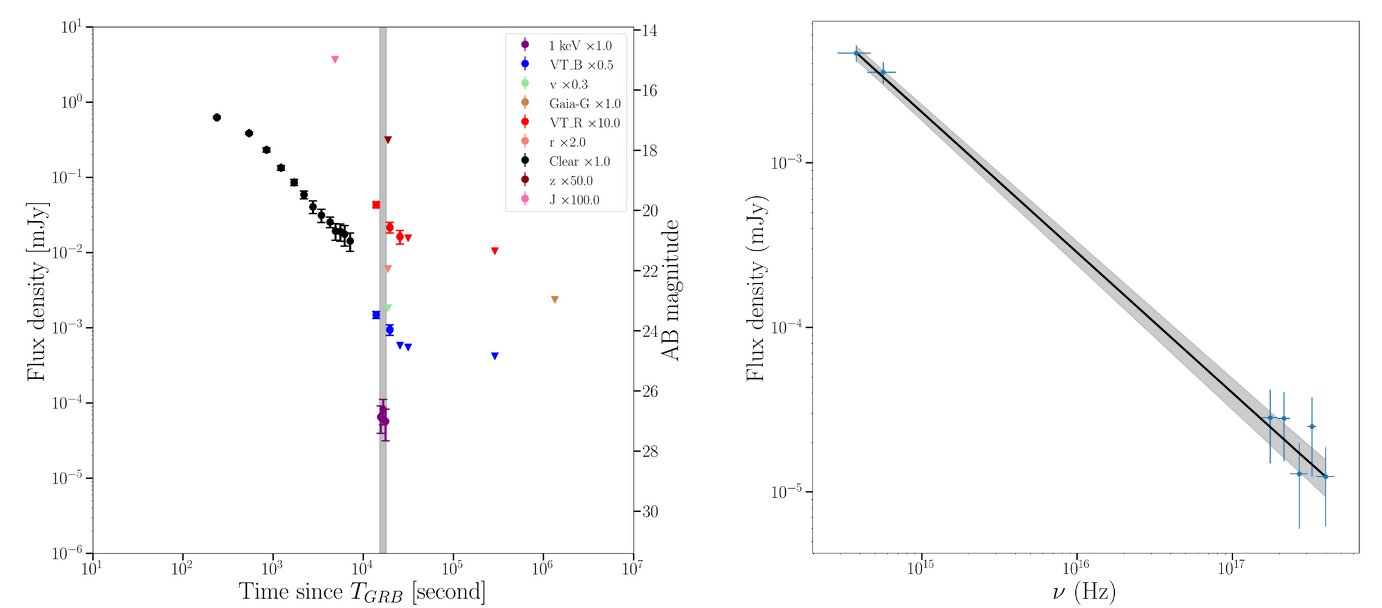}
    \caption{\textit{Left:} The X-ray to NIR observations of the GRB 241113B afterglow. Upper limits are denoted as downward triangles. The magnitudes have been corrected for the Milky Way visual extinction in the GRB line of sight using the reddening value $E(B-V) = 0.0784$ mag extracted from the \cite{Schlafly2011} dust map. The gray shaded area represents the time interval $T-T_0$ = [15,101--18,119~s] at which we extract the afterglow SED. \textit{Right:} The X-ray to NIR SED taken at $T-T_0$ = [15,101--18,119~s]. A simple power-law model is fitted to the data with a spectral slope $\beta = 0.83\pm0.04$.}
    \label{fig:Afterglow}
\end{figure*}

In Figure \ref{fig:Afterglow}, we show the complete X-ray to NIR afterglow light curve and spectral energy distribution (SED) obtained for GRB 241113B. 

\textbf{SED analysis:} We built the afterglow SED at a common time between the \textit{EP}/FXT and \SVOM/VT observations. The X-ray spectrum was directly extracted from the spectral analysis described in Section \ref{sec:xaglow_spec_analysis} but grouped with a a larger SNR (5 counts per bin). The \SVOM/VT spectral flux and associated errors had to be interpolated within the time interval centered on the middle of the X-ray observation epoch. The optical data were corrected for Galactic absorption, but we neglected potential host-dust absorption (see Section \ref{sec:redshift_constraints}). We fit the SED with a simple power law and a broken power law. The best fit ($\chi^2_\nu = 0.42$) was given by the power-law model with a x-ray to optical spectral index $\beta_{OX} = 0.83 \pm 0.04$ and illustrated in Figure \ref{fig:Afterglow}. This suggests that the X-ray and optical/NIR emission belong to the same power-law branch of the synchrotron spectrum.


\textbf{Light-curve analysis:} We then initiated a fit of the multiwavelength afterglow with a simple phenomenological model that allows us to extract the main features of the spectro-temporal evolution of the afterglow. For this, we used a smoothly broken power law (SBPL) model,
\begin{equation}
    F_\nu(t) = F_{0} \times \left[\left(\frac{t}{t_b}\right)^{-\delta\alpha_1} + \left(\frac{t}{t_b}\right)^{-\delta\alpha_2}\right]^{-1/\delta}\times \nu^{-\beta_{\rm OX}}\, ,
    \label{equ:aglow_pheno_model}
\end{equation}
where $F_{0}$ is a flux-density normalization factor and $\delta$ is a factor that shapes the smooth transition between the two temporal slopes at the break time ($t_b$). $\alpha_1$ and $\alpha_2$ are the temporal indices respectively before and after $t_b$. The best-fit model produces a common X-ray to NIR fading slope $\alpha_2 = -1.40^{+0.03}_{-0.04}$. Because we lack data at earlier times than $T_{\rm obs} - t_{0} = 200$~s, we cannot significantly constrain the potential rising slope of the optical afterglow as well as the sharpness of the peak brightness. The peak-brightness time is then constrained within $t_b = 263^{+68}_{-91}$~s after $t_{0}$ that we tentatively attribute 
to the deceleration of the jet into the circumburst medium.

We also performed a fit with a double smoothly broken power law to evaluate if we could have a late-time break during the X-ray emission ($T-t_0\approx 10^4$~s), but it led to an unconstrained model at late times and did not improve the fit at earlier times. 

Note that, contrary to our SED analysis, we found $\beta_{\rm OX} = 0.66^{+0.04}_{-0.03}$, but this value has to be taken with caution because the model expressed in Eq. \ref{equ:aglow_pheno_model} does not track the true spectral evolution of the afterglow synchrotron emission (and especially the potential presence of cooling breaks).
In Appendix \ref{appendix:pheno_fit}, we show the best-fit model with the 68$\%$ confidence level contours as well as the posterior distribution of the model parameters.


From these phenomenological fits, we tested the closure relations between $\alpha$ and $\beta$ in the framework of the standard fireball forward-shock model and in the simple top-hat jet structure \citep{Granot_Sari2002}. We considered that the forward-shock x-ray and optical synchrotron emission evolves in the so-called slow cooling regime ($\nu_m<\nu_{\rm opt} < \nu_{\rm X} <\nu_c$) where $\nu_m$ and $\nu_c$ are the characteristic synchrotron frequencies of the minimum-energy electrons and of the electrons cooling on the dynamical timescale, respectively. This gives us the relations:
\begin{equation}
    \begin{split}
        \mathrm{(ISM)~~} \alpha = 3(1-p)/4 \mathrm{~~and~~}\beta = (1-p)/2\, ,\\
        \mathrm{(Wind)~} \alpha = (1-3p)/4 \mathrm{~~and~~}\beta = (1-p)/2\, ,
    \end{split}
    \label{eq: closure relation}
\end{equation}
where $p$ is the index of the shock-accelerated electron power-law energy distribution.

This yields $\alpha_{\rm ISM} = $ $3\beta/2$ in the ISM constant-medium case and $\alpha_{\rm Wind} = (3\beta+1)/2$ in the wind medium case. Using $\beta = 0.66^{+0.04}_{-0.03}$ obtained from the multi-wavelength light curve analysis, we find $\alpha_{\rm ISM} = 0.99\pm0.06$ and $\alpha_{\rm Wind} = 1.49\pm0.06$. This simple approach suggests that the wind environment better reproduces the spectro-temporal evolution of the external shocks with an electron power-law energy distribution index $p = 2.31\pm0.08$. To confirm this quick diagnosis, we also performed a deeper investigation with a physical model.

\subsection{Afterglow Modeling with a Physical Approach}
\label{sec: afterglow modeling}
We performed physical modeling of the multiwavelength afterglow emission of GRB 241113B using the \textsc{VegasAfterglow} code \citep{Wang2026, Zhang2018}.\textsc{VegasAfterglow} is a C++ and Python-based framework for modeling GRB afterglow forward- and reverse-shock dynamics from early to late times, including energy-injection processes. It supports various jet structures, opening- and viewing-angle configurations, and multiple circumburst environment profiles (ISM, Wind, and custom). The code is computationally efficient and includes a built-in Bayesian inference framework through \textsc{Bilby} \citep{bilby_paper, bilby_doi}.

Given the sparse temporal sampling of the afterglow light curve and the absence of a spectroscopic redshift measurement, we adopted a simplified physical model to avoid overfitting. Specifically, we used a top-hat jet model dominated only by forward-shock dynamics, and tested two circumburst density profiles: a constant-density interstellar medium (ISM) and a stellar-wind environment (Wind) produced by the progenitor star prior to the GRB explosion. Several parameters were fixed based on physical arguments and preliminary fitting runs:

\begin{enumerate}
    \item We assumed that all electrons crossing the shock 
    discontinuity are accelerated and contribute to the synchrotron 
    radiation in the slow cooling regime ($\xi_e = 1$).
    \item The early optical afterglow rises rapidly and settles onto 
    a steady power-law decay within a few minutes, suggesting an 
    on-axis viewing geometry. This was confirmed by preliminary 
    fitting runs, which systematically favored a viewing angle, $\theta_v \approx 0^\circ$ 
    independently of the medium, microphysics parameters, or the redshift. 
    We therefore fixed $\theta_v = 0^\circ$ to reduce the number of 
    free parameters. Note that we later discuss the inconsistency between our data and the the off-axis scenario in Section \ref{sec:discussion}.
    \item As the phenomenological analysis does not reveal a jet break 
    at $T_\mathrm{jet,break} - T_\mathrm{GRB} < 10^4$\,s, we fixed 
    the jet core opening angle to a large value ($\theta_c = 20^\circ$) 
    to avoid introducing an artificial early break in the model 
    light curve.
\end{enumerate}

Since no spectroscopic redshift is available for GRB~241113B, we  ran the Bayesian inference on a grid of redshift values $z \in [0.5, 3.5]$ with a step of $\Delta z = 0.5$, fitting the ISM and Wind models independently at each redshift. The parameter inference was performed using the Dynamic Nested Sampling package \textsc{Dynesty} \citep{Speagle2020, Koposov2025}, which computes Bayesian posteriors and evidence simultaneously. We initialized the sampler with 500 live points and adopted a stopping criterion of $|\delta \ln\mathcal{Z}| = 0.5$. The log-evidence $\ln\mathcal{Z}$ is used as a Bayes factor to compare models, both across density profiles (ISM vs.\ Wind) and across redshifts. Model selection follows the Jeffreys scale \citep{Jeffreys1961, Kass1995}: $|\Delta \ln\mathcal{Z}| = 
|\ln\mathcal{Z}_1 - \ln\mathcal{Z}_2| > 5$ is taken as very strong evidence in favor of the model with the higher $\ln\mathcal{Z}$. The adopted priors on the free parameters are summarized in Table~\ref{tab:vegasafterglow_results}.

The best-fit model light curves and posterior distributions are shown in Appendix~\ref{appendix:vegasafterglow_fit} for both density profiles. A summary of the results is provided in Table~\ref{tab:vegasafterglow_results}.

\begin{table*}[h]
\centering
\renewcommand{\arraystretch}{1.5} 
\setlength{\tabcolsep}{2pt} 
\footnotesize

\begin{tabular}{lllccccccc}
\hline
Parameter & Scale & Prior & \multicolumn{7}{c}{ISM (with free $\theta_c$)} \\ \cmidrule(lr){4-10} 
          &       &       & z = 0.5 & z = 1.0 & z = 1.5 & z = 2.0 & z = 2.5 & z = 3.0 & z = 3.5\\
\hline
\hline
$\mathrm{E_{k,iso}}$ (erg) & $\mathrm{log_{10}}$ & [50 , 55] & $52.09^{+0.48}_{-0.92}$ & $52.65^{+0.32}_{-1.43}$ & $52.84^{+0.32}_{-1.84}$ & $52.96^{+0.39}_{-2.03}$ & $52.96^{+0.58}_{-2.05}$ & $53.20^{+1.19}_{-2.10}$ & $53.67^{+0.94}_{-2.39}$  \\
$\mathrm{\Gamma_0}$ & $\mathrm{log_{10}}$ & [1 , 4] & $3.19^{+0.21}_{-0.42}$ & $3.25^{+0.16}_{-0.36}$ & $3.35^{+0.08}_{-0.38}$ & $3.38^{+0.05}_{-0.34}$ & $3.42^{+0.06}_{-0.27}$ & $3.49^{+0.35}_{-0.18}$ & $3.70^{+0.22}_{-0.30}$  \\
p  & linear & [2.0 , 3.0]  & $2.36^{+0.07}_{-0.07}$ & $2.37^{+0.06}_{-0.07}$ & $2.40^{+0.07}_{-0.07}$ & $2.43^{+0.07}_{-0.05}$ & $2.51^{+0.26}_{-0.06}$ & $2.79^{+0.08}_{-0.21}$ &$2.83^{+0.05}_{-0.05}$  \\
$\mathrm{n_{ISM}}$ ($\mathrm{cm^{-3}}$) & $\mathrm{log_{10}}$ & [$-$3 , 2] & $1.11^{+0.57}_{-0.57}$ & $1.51^{+0.30}_{-0.38}$ & $1.71^{+0.19}_{-0.29}$ & $1.83^{+0.12}_{-0.20}$ & $1.82^{+0.13}_{-0.69}$ & $0.86^{+1.00}_{-1.95}$ & $-0.27^{+1.25}_{-0.93}$  \\
$\mathrm{\epsilon_{e}}$ & $\mathrm{log_{10}}$ & [$-$5 , 0.3] & $-1.29^{+0.19}_{-0.32}$ & $-1.35^{+0.13}_{-0.16}$ & $-1.39^{+0.09}_{-0.11}$ & $-1.43^{+0.07}_{-0.09}$ & $-1.50^{+0.08}_{-0.16}$ & $-1.71^{+0.18}_{-0.34}$ & $-1.93^{+0.21}_{-0.22}$ \\
$\mathrm{\epsilon_{B}}$ & $\mathrm{log_{10}}$ & [$-$5 , $-$1] & $-4.25^{+0.57}_{-0.49}$ & $-4.54^{+0.34}_{-0.31}$ & $-4.71^{+0.24}_{-0.18}$ & $-4.83^{+0.15}_{-0.11}$ & $-4.88^{+0.20}_{-0.09}$ & $-4.72^{+0.93}_{-0.22}$ & $-4.42^{+0.80}_{-0.45}$  \\
$\mathrm{\xi_{e}}$ &  & 1.0 & fixed   \\
$\mathrm{\theta_{v}}$ (rad) & & 0.0 & fixed  \\
$\mathrm{\theta_c}$ (rad) & linear & [0.04, 0.35] & $0.05^{+0.02}_{-0.01}$ & $0.05^{+0.01}_{-0.01}$ & $0.044^{+0.004}_{-0.003}$ & $0.042^{+0.003}_{-0.002}$ & $0.042^{+0.010}_{-0.001}$ & $0.05^{+0.11}_{-0.01}$ & $0.09^{+0.11}_{-0.04}$ \\
\hline
$\ln\mathcal{Z}$ & &  & $-25.4\pm0.3$ & $-27.2\pm0.3$ & $-29.1\pm0.3$ & $-31.7\pm0.4$ & $-32.9\pm0.3$ & $-33.9\pm0.3$ & $-34.5\pm0.3$ \\

\hline
\hline
Parameter & Scale & Prior & \multicolumn{7}{c}{Wind} \\ \cmidrule(lr){4-10} 
          &       &       & z = 0.5 & z = 1.0 & z = 1.5 & z = 2.0 & z = 2.5 & z = 3.0 & z = 3.5\\
\hline
\hline
$\mathrm{E_{k,iso}}$ (erg) & $\mathrm{log_{10}}$ & [50 , 55] & $52.64^{+1.52}_{-1.36}$ & $52.64^{+1.56}_{-0.95}$ & $52.78^{+1.36}_{-0.88}$ & $52.94^{+1.33}_{-1.02}$ & $52.91^{+1.32}_{-1.03}$ & $53.09^{+1.19}_{-1.50}$ & $53.10^{+1.21}_{-1.70}$  \\
$\mathrm{\Gamma_0}$ & $\mathrm{log_{10}}$ & [1 , 4] & $2.70^{+0.62}_{-0.60}$ & $2.79^{+0.84}_{-0.77}$ & $2.80^{+0.62}_{-0.57}$ & $2.86^{+0.55}_{-0.53}$ & $2.90^{+0.57}_{-0.52}$ & $2.94^{+0.53}_{-0.52}$ & $2.97^{+0.55}_{-0.49}$ \\
p  & linear & [2.0 , 3.0]  & $2.30^{+0.04}_{-0.04}$ & $2.29^{+0.04}_{-0.04}$ & $2.28^{+0.04}_{-0.05}$ & $2.28^{+0.04}_{-0.05}$ & $2.28^{+0.04}_{-0.05}$ & $2.27^{+0.05}_{-0.06}$ & $2.27^{+0.04}_{-0.05}$  \\
$\mathrm{A^*}$ & $\mathrm{log_{10}}$ & [$-$5 , 1] & $-2.63^{+0.80}_{-0.84}$  & $-2.23^{+0.84}_{-0.77}$ & $-2.00^{+0.89}_{-0.78}$ & $-1.85^{+0.81}_{-0.78}$ & $-1.78^{+0.87}_{-0.81}$ & $-1.67^{+0.74}_{-0.78}$ & $-1.68^{+0.85}_{-0.80}$ \\
$\mathrm{\epsilon_{e}}$ & $\mathrm{log_{10}}$ & [$-$5 , 0.3] & $-1.07^{+0.43}_{-0.43}$ & $-1.16^{+0.48}_{-0.43}$ & $-1.21^{+0.49}_{-0.42}$ &  $-1.25^{+0.44}_{-0.42}$  & $-1.29^{+0.48}_{-0.43}$ & $-1.31^{+0.46}_{-0.41}$ & $-1.35^{+0.47}_{-0.43}$ \\
$\mathrm{\epsilon_{B}}$ & $\mathrm{log_{10}}$ & [$-$5 , $-$1] &  $-2.87^{+1.22}_{-1.30}$ & $-2.95^{+1.28}_{-1.29}$ &  $-2.99^{+1.25}_{-1.20}$ & $-3.04^{+1.33}_{-1.35}$ &  $-3.10^{+1.38}_{-1.20}$ & $-3.16^{+1.29}_{-1.20}$ & $-3.21^{+1.39}_{-1.15}$ \\
$\mathrm{\xi_{e}}$ &  & 1.0 & fixed   \\
$\mathrm{\theta_{v}}$ (rad) & & 0.0 & fixed  \\
$\mathrm{\theta_c}$ (rad) & & 0.35 & fixed   \\
\hline
$\ln\mathcal{Z}$ & &  & $-15.0\pm0.3$ & $-15.2\pm0.3$ & $-15.0\pm0.3$ & $-15.2\pm0.3$ & $-15.1\pm0.3$ & $-15.1\pm0.3$& $-15.2\pm0.3$ \\
\hline
\hline

\end{tabular}%
\caption{Result of the Bayesian inference analysis of the GRB 241113B afterglow light curve using the \textsc{VegasAfterglow} code. The priors are uniformly distributed within their boundaries. The errors on the fitted parameters are given at the 68$\%$ confidence level.}
\label{tab:vegasafterglow_results}
\end{table*}

\subsubsection{ISM or Wind-like Profile?}

We investigate whether a constant-density ISM or a wind-like environment is preferred to describe the jet interaction of GRB 241113B.

\textbf{ISM model:}
The ISM scenario struggles to reproduce the multiwavelength afterglow, systematically underpredicting the X-ray flux density while overpredicting the NIR WINTER upper limit. The main sources of tension are described below:

\begin{enumerate}
\item In the ISM case, the coasting phase produces a broad peak in the light curve, unlike the flatter early-time behavior expected in a Wind environment. Consequently, the model is highly sensitive to the sparsely sampled early optical data ($T - T_{\rm GRB} < 500$ s). For example, the rest-frame deceleration time, $t_{\rm dec,z} = t_{\rm dec}/(1+z)$, scales as $\Gamma_0^{(8-2s)/(3-s)}$ (with $s=0$; see Eq. \ref{eq:gamma_0}). The Bayesian inference therefore favors unrealistically large values of $\Gamma_0 > 10^3$ to match the potential observed peak.

\item The ISM model cannot simultaneously reproduce the optical and X-ray temporal slopes while satisfying the closure relations described in Section~\ref{sec:pheno_fit} for reasonable electron indices $p \in [2.0, 3.0]$. Acceptable values of $p$ ($p\approx 2.4$--2.5) are recovered only by introducing a jet break at $T - T_{\rm GRB} \approx 8 \times 10^{3}$~s and relaxing the constraint on the jet opening angle $\theta_c$. This solution implies narrow jets ($\theta_c \approx 3^\circ$–$4^\circ$) but still fails to reproduce the data at high redshift ($z > 2$), where $\Gamma_0$, $E_{k,\rm iso}$, and $n_{\rm ism}$ reach extreme values.

\end{enumerate}

\textbf{Wind-like model:}
The Wind-like environment is less sensitive to early-time sampling and predicts intrinsically steeper forward-shock slopes than the ISM case. We find that it naturally reproduces the observed single power-law behavior across the NIR/optical and X-ray bands at all tested redshifts, without relaxing the constraint on $\theta_c$. \\

The inferred electron index (independent of the redshift value) is $p \approx 2.3$ (Table~\ref{tab:vegasafterglow_results}), corresponding to $\alpha_{\rm OX} \approx 1.48$ and $\beta_{\rm OX} \approx 0.65$, in good agreement with the phenomenological fit (Section~\ref{sec:pheno_fit}). In this framework, both optical and X-ray emission lie on the same synchrotron segment in the slow-cooling regime below the cooling frequency: $\nu_m < \nu_{\rm opt} < \nu_{\rm X} < \nu_c$ (Figure~\ref{fig:Afterglow}).

In summary, the ISM model requires an additional late-time break relative to the phenomenological fit in order to reproduce both the optical and X-ray emission for redshifts $z\lesssim2.0$. This entails a small jet opening angle ($\theta_c\approx 3^\circ$--$4^\circ$ to satisfy the pre–jet-break closure relations. However, even after relaxing the constraints on $\theta_c$, for high redshifts ($z>2$), the data and model discrepancies remain irreconcilable and the ISM scenario still fails to account for the broadband afterglow. We thus favor a Wind-like environment surrounding GRB 241113B because it (1) naturally explains the steep optical-to-X-ray slope without invoking a late-time break (that cannot be reasonably constrained by the dataset), (2) provides better fits (based on the $|\Delta \ln\mathcal{Z}|$ criterion) than the ISM model (with fixed or relaxed $\theta_c$), and (3) yields more physically appropriate values of $E_{k,\rm iso}$ and $\Gamma_0$ compared to the ISM model with relaxed $\theta_c$ values.

\subsubsection{Constraints on the Jet's Bulk Lorentz Factor, $\Gamma_0$}
\label{sec: bulk_lorentz_factor}
In the framework of the standard fireball picture, the relativistic jet first expands in a coasting phase before sweeping up enough material and transitioning to a self-similar deceleration phase at a time $t_{\rm dec}$ \citep{Piran1999,Kumar_Zhang2015}. During the coasting phase, the jet will reach its maximum bulk Lorentz factor $\Gamma_0$ before gradually decelerating. Many models were proposed in the past to describe the $\Gamma(t)$ evolution and the transition between the coasting and deceleration phase; see \citet{Ghirlanda2018}. The peak time of the afterglow in the rest frame is actually connected to the blast-wave dynamics and parameterized as

\begin{equation}
    \Gamma_0 = k \left(\frac{E_0}{n_0m_pc^{5-s}}\right)^{\frac{1}{8-2s}} \times t_{{\rm dec},z}^{-(3-s)/(8-2s)}\, ,
    \label{eq:gamma_0}
\end{equation}
where $k$ is a normalization factor that depends on the model assumptions about the transition from the coasting to the deceleration  phase \citep[see Section 5 of ][]{Ghirlanda2018}, $E_0$ is the initial isotropic kinetic energy of the fireball, $m_p$ is the proton's mass, $n_0$ is the normalization factor of the circumburst medium density profile $n(R) = n_0\, r^{-s}$, $t_{{\rm dec},z} = t_{\rm dec}/(1+z)$ is the rest-frame deceleration time, and $s = 0$ or 2 for a constant-density or a wind-like circumburst medium, respectively. Assuming that the estimated peak time of the optical afterglow ($t_b = 263^{+68}_{-91}$ s; see Section \ref{sec:afterglow_analysis}) corresponds to the observed deceleration time $t_{\rm dec}$, and using the fiducial values for the different parameters, $E_0 = E_{\gamma,{\rm iso}}/\eta_\gamma$ erg with $\eta_\gamma = 20\%$, the typical value of radiative efficiency of the prompt emission \citep{Kumar_1999,Nava2014}, and $n_0 = 1~{\rm cm}^{-3}$ (in the ISM case, $s = 0$) or $n_0 \approx 3\times 10^{35}\times A_*~{\rm cm}^{-1}$ (in the Wind case, $s = 2$), with $A_* = (\dot{M}/10^{-5}) \times (v_{w}/10^{3})^{-1}$ where $\dot{M}$ is the mass-loss rate in solar masses per year and $v_{w}$ is the speed of the stellar wind in $\mathrm{km~s^{-1}}$ \citep{Chevalier_Li_2000}, we obtain
\begin{itemize}
    \item ISM: $\Gamma_0 [z=0.5\rightarrow3.5,~t_{\rm dec} = 263~{\rm s}]  \approx k\times [197 \rightarrow 297]$;
    \item Wind: $\Gamma_0 [z=0.5\rightarrow3.5,~t_{\rm dec} = 263~{\rm s}] \approx k\times [121 \rightarrow 159]$  .
\end{itemize}

From our \textsc{VegasAfterglow} fit, while we do not tightly constrain $\Gamma_0$, we observed that we converged toward high values for the lower limits such as $\Gamma_{0,{\rm min}} (z=0.5 - 3.5) \approx [125 - 263]$ in the wind medium case. 

First, we conclude that our different $\Gamma_0$ constraints are in good agreement with the expected ultrarelativistic regime associated with GRB jets, in particular to avoid a strong $\gamma$-ray opacity due to pair production \citep{Ruderman1975}. Second, this analysis reinforces our idea that the measured $t_{b} = 263$~s likely corresponds to the observed deceleration phase of the GRB 241113B relativistic jet.

\subsubsection{Constraints on the Prompt Radiative Emission Mechanism}

From the posterior distributions of $E_{k,{\rm iso}}$ and $E_{\gamma,{\rm iso}}$, we can derive fundamental quantities of the prompt emission such as the radiative efficiency of the dissipative mechanism during the prompt phase, $\eta_\gamma$, as  \citep{Zhang2007}:

\begin{equation}
    \eta_\gamma = \frac{E_{\gamma,{\rm iso}}}{E_{\gamma,{\rm iso}}+E_{k,{\rm iso}}}\, .
\end{equation}
\noindent
The distribution of $\eta_\gamma$ values in GRB jets has long been debated in the literature, induced by the difficulty to characterize the nature of the jet composition and its energy distribution, as well as the mechanism responsible for the internal kinetic energy dissipation into radiation \citep[see, ][]{Beniamini_Nava_Piran_2016}. On the one hand, the internal-shock model usually infers low radiative efficiencies with $\eta_\gamma \lesssim 20\%$ \citep{Kobayashi_Piran_Sari_1997,Daigne_Mochkovitch_1998,Kumar_1999,Nava2014}, even if in some particular conditions $\eta_\gamma$ may reach high values from 50$\%$ \citep{Daigne_Bosnjak_Dubus_2011} up to nearly unity \citep{Beloborodov_2000}. On the other hand, GRB prompt magnetic reconnection models (such as ICMART) naturally allow for much higher $\eta_\gamma\approx 50$\%--90\%  \citep{Zhang_Yan_2011}.

In our case, we can derive the posterior distributions of $\eta_\gamma$ from the posterior distributions of our \textsc{VegasAfterglow} fits with the Wind-like medium; see Table \ref{tab:posterior_radiative_efficiency}.

\begin{table}[]
    \centering
    \renewcommand{\arraystretch}{1.4} 
    \begin{tabular}{c|c|c|c}
        \toprule
        z & $E_{\gamma,{\rm iso}}$ & $E_{k,{\rm iso}}$ &  $\eta_\gamma$ \\
          & log$_{10}$(erg) & log$_{10}$(erg) &   \\
        \midrule
        0.5 & $50.92^{+0.08}_{-0.07}$ & $52.64^{+1.52}_{-1.36}$ & $0.02^{+0.29}_{-0.02}$ \\
        1.0 & $51.54^{+0.08}_{-0.07}$ & $52.64^{+1.56}_{-0.95}$ & $0.06^{+0.36}_{-0.05}$ \\
        2.0 & $52.10^{+0.09}_{-0.07}$ & $52.94^{+1.33}_{-1.02}$ & $0.13^{+0.44}_{-0.12}$ \\
        3.0 & $52.40^{+0.08}_{-0.07}$ & $53.09^{+1.19}_{-1.50}$ & $0.17^{+0.69}_{-0.15}$ \\
        3.5 & $52.51^{+0.09}_{-0.07}$ & $53.10^{+1.21}_{-1.70}$ & $0.18^{+0.73}_{-0.17}$ \\
        \bottomrule
    \end{tabular}
    \caption{Constraints on the radiative efficiency, $\eta_\gamma$, of the prompt emission to convert kinetic energy into $\gamma$-rays as function of the GRB 241113B redshift.}
    \label{tab:posterior_radiative_efficiency}
\end{table}

In a general manner, the prompt radiative efficiency of GRB~241113B is not well constrained mostly because of the uncertainties on $E_{k,{\rm iso}}$ from the afterglow modeling. This may be due to the fact that the X-ray synchrotron photons are affected by  significant Comptonization (not taken into account in the afterglow modeling) or lie below the synchrotron cooling break frequency, and thus they may weakly track the kinetic energy, $E_{k,{\rm iso}}$ of the jet post prompt emission \citep{Nava2014,Beniamini_Nava_Piran_2016}. For $z\leq1.0$, we tentatively infer a maximum value $\eta_\gamma \lesssim$ 40$\%$. At higher redshifts, we cannot bring any meaningful constraints on $\eta_\gamma$. 

%% file: Section/6_Discussion.tex
\section{Discussion}
\label{sec:discussion}

\subsection{Sakamoto Subclassification}
\label{sec: sakamoto discussion}
\begin{figure*}[htbp]
     \centering
     \begin{subfigure}[t]{0.48\textwidth}
         \centering
         \includegraphics[width=\linewidth]{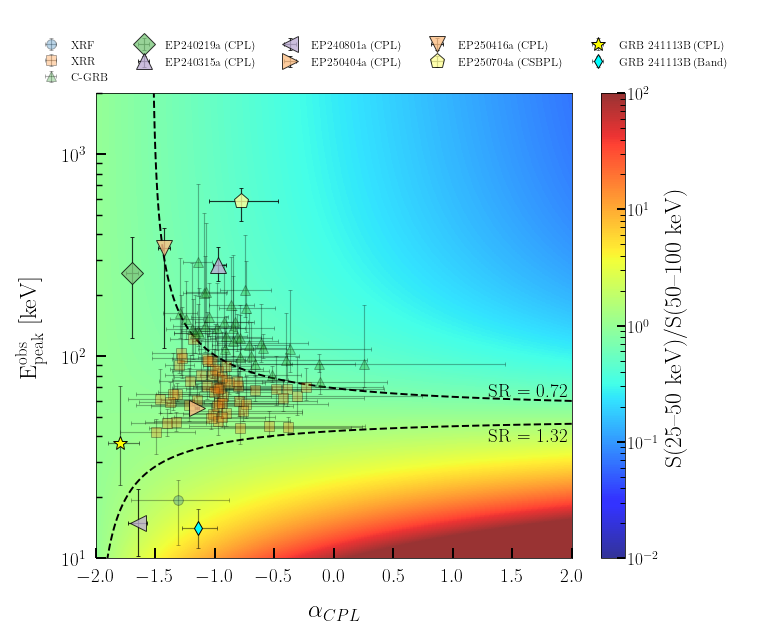}
         \caption{GRB 241113B  compared with other GRBs best fitted by a CPL from BAT3 \citep{Lien_2016}. This figure corresponds to the parametric plane of a CPL: the $E^{\rm obs}_{\rm peak}$ as a function of $\alpha_{\rm CPL}$. The color map corresponds to the fluence ratio defined by \cite{Sakamoto2008} to classify XRFs/XRRs/C-GRBs. The black dashed lines correspond to separations between subclasses of Type II GRBs as defined by \cite{Sakamoto2008}; GRBs lying below the line SR $=1.32$ are classified as XRFs, those above the line of SR $=0.72$ are C-GRBs, and  the remainder (enclosed by the two black lines) are  XRRs. In this plot GRB 241113B is clearly classified as a XRR.}
         \label{fig:cpl_bat}
     \end{subfigure}
     \hfill 
     \begin{subfigure}[t]{0.48\textwidth}
         \centering
         \includegraphics[width=\linewidth]{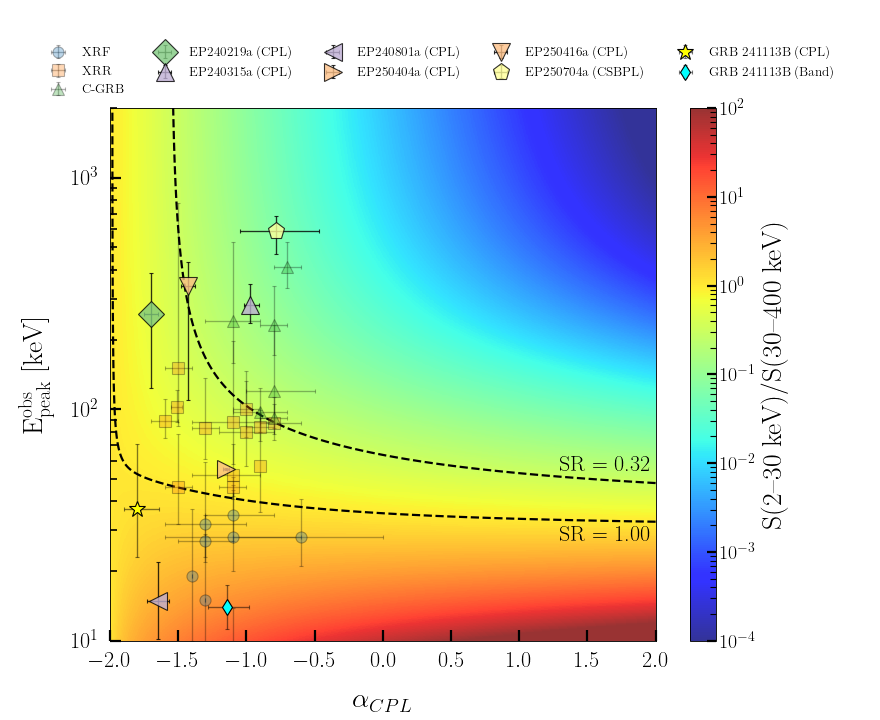}
         \caption{GRB 241113B compared with other GRBs best fitted by a CPL from HETE-2 \citep{Sakamoto2005}. This figure corresponds to the parametric plane of a CPL:  the $E^{\rm obs}_{\rm peak}$ as a function of $\alpha_{\rm CPL}$. The color map corresponds to the fluence ratio defined by \cite{Sakamoto2005} to classify XRFs/XRRs/C-GRBs. The black dashed lines correspond to separations  between subclasses of Type II GRBs as defined by \cite{Sakamoto2005}; GRBs lying below the line SR $=1.00$ are classified as XRFs, those above the line of SR $=0.32$ are C-GRBs, and the remainder (enclosed by the two black lines) are XRRs. In this plot GRB 241113B can be classified as a XRF or marginally as a XRR.}
         \label{fig:cpl_hete-2}
     \end{subfigure}
     
     \caption{GRB 241113B compared with a sample of GRBs  best fitted by a CPL (\textit{a}: BAT3, \citealp{Lien_2016}; \textit{b}: HETE-2, \citealp{Sakamoto2005}). They show the parametric plane of a CPL: the $E^{\rm obs}_{\rm peak}$ as a function of $\alpha_{\rm CPL}$, plus a color map corresponding to the fluence ratio used to classify Type II bursts (\textit{a}: \citealp{Sakamoto2008}; \textit{b}: \citealp{Sakamoto2005}). EP240219a \citep{Yin_triggering}, EP240315a \citep{Liu_2025}, EP240801a \citep{Jiang_2025}, EP250404a \citep{Yin2}, EP250416a \citep{Zhao_2026} and EP250704a \citep{Li_2026} are also shown for comparison.}
     \label{fig:Sakamoto CPL}
\end{figure*}

Under the hypothesis that GRB 24113B is a type II GRB (see section \ref{sec: Result of spectral analysis}), we explored its classification as an XRF/XRR following the subclassification of \cite{Sakamoto2005,Sakamoto2008}.As pointed out in Section \ref{sec: Result of spectral analysis}, GRB 241113B is best fitted by a CPL and can be classified as either an XRR or XRF using the fluence ratio from \cite{Sakamoto2008} or \cite{Sakamoto2005}, respectively. This change in classification of this soft burst depending on the chosen fluence ratio is a smoking-gun signature of the lack of consistency between the two formulations of the Sakamoto subclassification, contrary to  Section 2 of \cite{Sakamoto2008}. Indeed, when looking at the parametric plane for a CPL and the evolution of the fluence ratio (Figure \ref{fig:Sakamoto CPL}), one can see that the separation lines between the subclasses (represented by black dashed lines) diverge in the case of \cite{Sakamoto2008}, while they define a closed area with the \cite{Sakamoto2005} fluence ratio. 

This divergence of the \cite{Sakamoto2008} fluence ratio, in the context of a CPL, was already discussed by \cite{Yin_triggering}; it means that for a given $\alpha$ ($\alpha
\lesssim -1.5$), all bursts, independent of their $E^{\rm obs}_{\rm peak}$, end up classified as XRRs. The behavior contradicts the purpose of the fluence ratio, that aims to differentiate soft GRBs (low $E^{obs}_{\rm peak}$) from hard GRBs (high $E^{\rm obs}_{\rm peak}$). The consequence of the fluence ratio of \cite{Sakamoto2008} is that the XRR population is overestimated. The fluence ratio on a larger band, from \cite{Sakamoto2005}, defines a closed area that follows better the burst's $E^{\rm obs}_{\rm peak}$. Hence, from Figure \ref{fig:Sakamoto CPL}, one can say that \cite{Sakamoto2005} offer a better criterion to subclassify soft GRBs than the one provided by \cite{Sakamoto2008}. Ultimately, this demonstrates that observing the prompt emission of GRBs across a broad energy range can help distinguish between soft and hard Type II GRBs.

\subsection{Shock Breakout}
Most LL-GRBs are associated with ``shock breakout''  \citep[SBO;][]{Campana_2006,Soderberg_2008,Nakar_2012}. SBOs create bursts that do not arise from a relativistic jet, but rather owing to a shock (coming from the implosion of the star) that spreads to the outer layer and ``breaks out" of it  (producing more or less spherical, unbeamed emission). A potential jet due to a compact object forming inside the progenitor
can happen, but it is not powerful enough to drill through the star's outer shell, resulting in a ``choked jet'' \citep{Nakar_2012}. LL-GRBs are also usually associated with relatively nearby SNe (e.g., GRB 980425/SN1998bw, \citealp{Kaneko_2007}; GRB 031203/SN 2003lw, \citealp{Sazonov_2004}, \citealp{Malseani_2004}; GRB 100316D/SN 2010bh,  \citealp{Starling_2011}), some of which are clear XRFs (e.g., XRF 060218/SN 2006aj, \citealp{Campana_2006}; XRF 080109/SN 2008D, \citealp{Soderberg_2008}).

GRB 241113B has an isotropic $\gamma$-ray luminosity ($L_{\gamma,{\rm iso}}$) that ranges from 
$3.26^{+0.63}_{-0.42}\times10^{49}$ to 
$4.03^{+0.83}_{-0.58}\times10^{51}~{\rm erg}~{\rm s}^{-1}$ for $z=0.5$ and $z=3.5$, respectively. Hence, at low redshift $(z<1)$, GRB 241113B can be considered as an LL-GRB, and thus could be related to an SBO event. However, as stated by \cite{Nakar_2012}, SBOs should satisfy the relation 
\begin{equation}
t^{\rm obs}_{\rm bo}\approx 20~{\rm s} \left(\frac{E_{\rm bo}}{10^{46}~{\rm erg}}\right)^{1/2}\left(\frac{T_{\rm bo}}{50~{\rm keV}}\right)^{-(9+\sqrt{3})/4}\, ,
\label{eq:Nakar}
\end{equation}
where $t^{\rm obs}_{\rm bo}$ is the timescale of the SBO as seen by an observer, $E_{\rm bo}$ is the isotropic energy release by the SBO, and $T_{\rm bo}$ is the energy temperature of the SBO.  In the case of GRB 241113B, which has a significant redshift $(z\geq0.5)$, we have to scale Eq. \ref{eq:Nakar} as a function of the redshift. If we apply the notation used throughout this paper, we have
\begin{equation}
T^{\rm obs}_{90}\sim 20~{\rm s} \left(\frac{E_{\gamma,{\rm iso}}}{10^{46}~{\rm erg}}\right)^{1/2}\left(\frac{E^{\rm src}_{\rm peak}}{50~{\rm keV}}\right)^{-(9+\sqrt{3})/4}~(1+z)\, ,
\label{eq:Nakar_redshift}
\end{equation}
\noindent
where $E^{\rm src}_{\rm peak}$ is the energy peak in the source rest frame, i.e., $E^{\rm src}_{\rm peak} = (1+z) E^{\rm obs}_{\rm peak}$. Plugging the values listed in Table \ref{tab: joint_spectral_results} and Table \ref{tab:posterior_radiative_efficiency} into Eq. \ref{eq:Nakar_redshift}, we find that $t^{\rm obs}$ should be $\sim7000$~s and $\sim9000$~s at redshift 0.5 and 3.5, respectively. If we compare these values with the duration of the burst reported in Table \ref{tab:t90_comparison}, this is more than two orders of magnitude greater than the observed duration; such a difference is, we think, too strong to be due to the sensitivity limitation. Hence, we consider it very unlikely that GRB 241113B could be explained by a SBO. A relativistic jet should be invoked to describe this burst prompt emission and afterglow emission as demonstrated in \ref{sec:afterglow_analysis}.

\subsection{The Off-axis Model}
As discussed in Section \ref{sec:pheno_fit}, GRB 241113B exhibits an early peak ($t_{\text{b}} \approx 263$~s) bump followed by a normal decay without late-time rebrightening. This kind of optical afterglow is typical of GRBs \citep[45/146 events;][]{Li2012}, and can be considered as ``textbook versions of light curves as predicted by the standard fireball model'' \citep{Liang2013}. Indeed, GRB 241113B's  afterglow temporal decay and spectral index, $\alpha_{2} = -1.40^{+0.03}_{-0.04}$ and $\beta_{\text{\rm OX}} = 0.66^{+0.04}_{-0.03}$ (respectively), satisfy nicely the closure relation expected for the standard fireball model of a top-hat jet forward shock with synchrotron emission of electrons in a slow-cooling regime \citep[see][and closure relation summarized in Eq. \ref{eq: closure relation}]{Granot_Sari2002}. In particular, the temporal decay corresponds to the case of an electron spectral index of 2.2, which is close to the usual value observed for most GRBs \citep{Curran2010}. We can also add the fact that GRB 241113B follows 
the Amati relation (see Figure \ref{fig:amati_relation}), and is neither too soft nor too energetically weak. If GRB 241113B happened to be off-axis, one would have expected the burst to deviate from the aforementioned relation and from the standard afterglow scenario \citep{Chen_2026}. Hence, we are confident that, as shown by our afterglow fit (see Section \ref{sec: afterglow modeling}), the GRB 241113B viewing angle should be small; it might have been observed close to the jet axis.

Furthermore, the early peak of the afterglow ($t_{\text{b}} < 270$~s), together with its high brightness ($\sim 16.7$ mag), would be unusual for an off-axis GRB, and opposite of what is commonly observed for off-axis GRB afterglows, which mostly peak at $10^{3}$--$10^{4}$~s, and are rather faint, $\sim 18$~mag; see, e.g., the afterglow light curves of XRF 080330 \citep{Guidorzi2009}, XRF 030723  \citep{Butler_2005}, XRF 050406 \citep{Schady2006}, and XRR 050408 \citep{UgartePostigo2007}. Off-axis GRB 080710 was also very bright at its peak ($\sim16$ mag), but it is classified as a C-GRB by \citep{Kruhler2009}. According to \citet{Rossi2002} (see their Figs. 1 and 2), which details the off-axis scenario in the case of a universal jet, the afterglow should have a shallow decay, $\alpha \ge -0.8$, which is not as steep as the one we measure for GRB 241113B.

We tested the off-axis uniform jet model of \citet{Yamazaki2002} to check whether GRB 241113B is an GRB off-axis from the prompt emission point of view, applying the method of \cite{Butler_2005}. The model of \cite{Yamazaki2002} states that $\text{E}^{\text{src}}_{\text{peak}}\propto\delta^{-1}$ and $\text{E}_{\gamma,\text{iso}} \propto \delta^{-3}$, where $\delta$ is the Doppler factor, normalized with respect to an on-axis GRB ($\delta \approx [1+{\Gamma}^{2}(\theta_{\nu}-\Delta\theta)^{2}]$). However, instead of using fiducial values of the bulk Lorentz factor as done by \cite{Butler_2005}, we used the values of $\Gamma_{0}$ found in Section \ref{sec: bulk_lorentz_factor}. We then took as reference for $E_{\gamma,\text{iso}}$ and $E^{\text{src}}_{\text{peak}}$ the ones found for C-GRBs in \cite{Bi_2018}: $\log(E_{\gamma,{\rm iso}}) = 53.1^{+0.7}_{-1.0}$, $E^{\rm obs}_{\rm peak} = 257.7\pm14.1$~keV. In the off-axis scenario, C-GRBs should be by definition on-axis bursts, and these two quantities ($E_{\text{peak}}$, $E_{\gamma,\text{iso}}$) should vary proportionally to the normalized Doppler factor as mentioned before. Hence, we can express  $E_{\rm peak}$ and $E_{\gamma,{\rm iso}}$ as a function of $\theta_{\nu}-\Delta\theta$, and find that in order to recover the values observed for GRB 241113B, the separation between the two angles should not exceed $\sim3.80^\circ$. However, it is important to note that the lack of afterglow data means that $\Gamma_{0}$ is rather poorly constrained (see Table \ref{tab:vegasafterglow_results}); thus, significant uncertainties remain in this scenario.

\subsection{Dirty Fireball}
The dirty fireball model \citep{Huang2002} predicts that soft bursts arise from jets with $\Gamma \ll 100$. However, as computed in Section \ref{sec: bulk_lorentz_factor}, or from the afterglow fit, and if we assume that $\Gamma \approx \Gamma_{0}$, the bulk Lorentz factor of GRB 241113B seems to be greater than 100 in both the ISM and Wind scenarios. Since GRB 241113B follows the Amati relation, its jet bulk Lorentz factor is very unlikely to be weak; otherwise, it would lie below this relation \citep[EP241113a,][]{Dai2026}. Hence, we disfavor the dirty fireball model to explain the softness of GRB 241113B.

\subsection{GRB 241113B and FXTs}
Since its launch, \EP\ has detected numerous GRBs, XRFs/XRRs, and fast X-ray transients (FXTs). Contrary to GRBs, the progenitors of FXTs are not clearly known; they could be produced throughout many channels (see \citealp{Vasquez2023}, and reference therein). These transients are characterized by a low $E^{\rm obs}_{\rm peak}<5$~keV, with $\alpha < -2$ (significantly smaller than the typical photon index of XRFs, XRRs, and C-GRBs), and lie below the Amati relation (e.g. see Figure \ref{fig:amati_relation}). However, GRB 241113B does not show such extreme spectral softness and is much too high in the Amati relation to be classified as an FXT; it is more related to the soft tail of the distribution of GRBs.

%% file: Section/7_Conclusion.tex
\section{Conclusion}
\label{sec:conclusion}
GRB 241113B / EP241113b represents the first joint detection of a soft GRB by the {\it Einstein Probe} and {\it SVOM}. Our temporal and spectral analysis of the prompt emission indicates that GRB 241113B is consistent with a collapsar-driven Type II event. On the afterglow side, GRB 241113B exhibits a typical brightness and temporal behavior consistent with classical GRBs (C-GRBs) involving the presence of an ultrarelativistic jet (with $\Gamma_0 > 100$) likely observed on-axis. We also favor the hypothesis that the external shocks between the jet and the circumburst environment occurred in a wind-like medium produced by the stellar wind of the initial massive-star progenitor.
The simultaneous detection of this burst by {\it EP}/WXT and {\it SVOM}/ECLAIRs and GRM enables us to confirm the prompt $T_{90}$–energy correlation previously reported for BATSE and {\it Fermi}/GBM bursts.

Our prompt spectral analysis further classifies GRB 241113B as a X-ray rich (XRR) or X-ray flash (XRF) burst according to the Sakamoto subclassification, placing it at the faint end of the isotropic-equivalent gamma-ray energy ($E_{\gamma,\mathrm{iso}}$) distribution of typical Type II GRBs. In this context, we revisit the standard interpretation of the Sakamoto subclassification scheme. We tend to favor the one published by \citet{Sakamoto2005}, as that from \citet{Sakamoto2008} tends to overpredict the number of XRR GRBs. Also, the fluence ratio from \citet{Sakamoto2005} is better suited to the  observations made with {\it EP}/WXT and \SVOM/ECLAIRs with their low-energy threshold (at 0.5 and 4 keV, respectively), plus the extension of GRM up to 1 Mev. 

Current evidence suggests that GRB 241113B originates from the same class of transient events as energetic Type II C-GRBs. This supports a phenomenological continuity in GRBs that extends down to events detected at just a few tens of keV. Notably, because GRB 241113B's jet appears to have been observed on-axis with ultrarelativistic motion, we can rule out two common models typically used to explain soft GRBs: the dirty fireball and the off-axis perspective. Consequently, the softness of GRB 241113B is driven neither by geometric viewing effects nor by heavy baryon loading in the jet. Instead, this observation indicates that the burst softness is likely intrinsic in nature, potentially resulting from a jet with a low bulk Lorentz factor contrast between shells, photospheric emission, or the presence of a hot cocoon surrounding the jet.

Up to now, FXTs observed by {\it EP} have only upper limits on their $\gamma$-ray flux for soft spectrum template, in 10--1000~keV band, is $\sim2\times10^{-7}$~erg~cm$^{-2}$~s$^{-1}$ (for 1~s exposure) and $\sim7\times10^{-8}$~erg~cm$^{-2}$~s$^{-1}$ (for burst with more than 100~s durations; \citealt{Zhang_2025,Ravasio_2026}), whereas the ECLAIRs sensitivity, for spectrum in $E^{-2}$, is $4.2\times10^{-8}$~erg~cm$^{-2}$~s$^{-1}$ (for 1~s, in 4--120~keV), and $4.2\times10^{-9}$~erg~cm$^{-2}$~s$^{-1}$ (for 100~s, in 4--120~keV; \citealt{Dagoneau_2021}). Therefore, ECLAIRs could play a key role in unveiling the nature of these mysterious FXTs and contribute to the discovery of new soft transients. Such events are of particular interest because their energetics do not always appear to follow the Amati relation, potentially pointing to a distinct physical origin or emission mechanism \citep{Chen_2026}.

%% file: Section/8_Appendix.tex





\begin{appendix}
\onecolumn

\section{Detection significance in high-energy instruments}
\label{appendix: significance}

\begin{table}[h!]
    \centering
    \renewcommand{\arraystretch}{1.2} 
    \begin{tabular}{l c c} 
        \toprule
        \textbf{Instrument} & \textbf{Energy Band} & \textbf{Significance [$\sigma$]} \\
        \midrule
        
        WXT & 0.5--4 keV & 10.7 \\
        \midrule
        
        ECLAIRs & 4--30 keV & 27.0 \\
                & 4--120 keV & 26.9 \\
        \midrule

        \addlinespace 

        GRD1 & 4--30 keV & 8.3 \\
        GRD2 & 4--30 keV & 10.5 \\
        GRD3 & 4--30 keV & 4.2 \\
        GRM  & 4--30 keV & 12.9 \\
        
        
        \addlinespace 
        
        GRD1 & 50--300 keV & 3.3 \\
        GRD2 & 50--300 keV & 2.9 \\
        GRD3 & 50--300 keV & 2.2 \\
        GRM  & 50--300 keV & 4.8 \\


        
        \bottomrule
    \end{tabular}
    \caption{Significance of the GRB 241113B signal across all high-energy instruments 
    in different energy bands.}
    \label{tab: significance}
\end{table}

\FloatBarrier
\section{WXT Independent Fit}
\label{appendix: WXT}




\begin{table}[h]
\centering
\renewcommand{\arraystretch}{1.3} 
\begin{tabular}{l c c c c c}
\hline
Model & $N_{\rm H,X}$ & $\alpha$ & ${K_{1}}^{a}$ & ${F_{X}}^{b}$ & CSTAT/dof \\ 
& ($10^{22}$ cm$^{-2}$) & & (ph~cm$^{-2}$~s$^{-1}$) & (${10^{-9}}$ erg cm$^{-2}$ s$^{-1}$) & \\ 
\hline
PL & $0.74^{+0.36}_{-0.31}$ & $2.55^{+0.92}_{-0.82}$ & $2.58^{+2.52}_{-1.20}$ & $2.98^{+0.75}_{-0.59}$ & 17/19 \\ 
\hline
\end{tabular}

\vspace{1ex}
\parbox{1.0\textwidth}{
    \footnotesize
    $^a$ Normalization at 1 keV. \\
    $^b$ Flux in 0.5--4.0 keV in erg/cm$^2$/s. \\ 
}

\caption{{\it EP}/WXT independent fit. Here all errors are given at the 68\% C.I.}
\label{tab:WXT}
\end{table}

\FloatBarrier
\newpage
\section{Joint fit SED }
\label{appendix: Joint SED}

\begin{figure}[h!]
    \centering
    \includegraphics[width=0.60\textwidth]{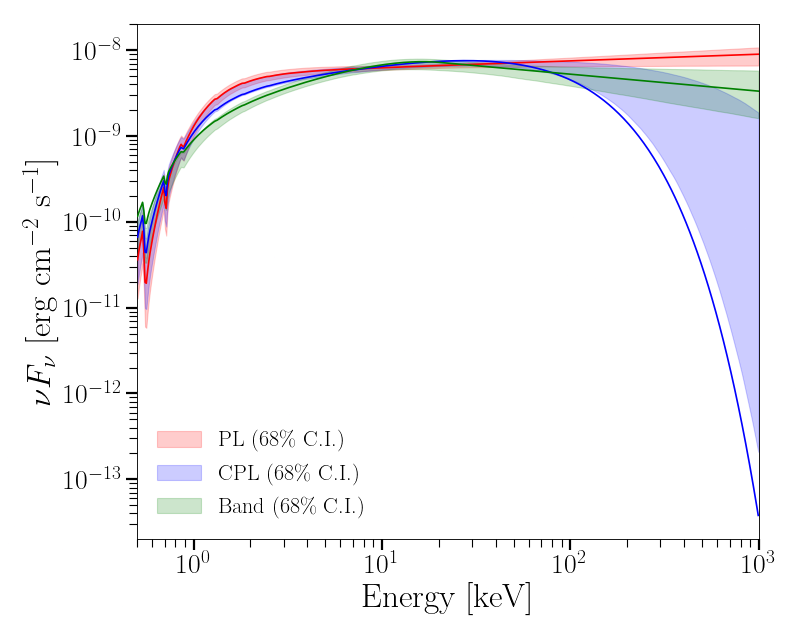}
    \caption{GRB 241113B prompt emission SED for PL (in red), CPL (in blue) and Band (in green) models, over 0.5--1000~keV. The continuous line corresponds to the best fit for each model, while the colored shadows are the 68\%~C.I. (1~$\sigma$) envelop.}
    \label{fig:theo_spec}
\end{figure}

\FloatBarrier
\section{EDH Classification}
\label{appendix: EDH classification}
\begin{figure}[h]
     \centering
     \includegraphics[width=0.60\textwidth]{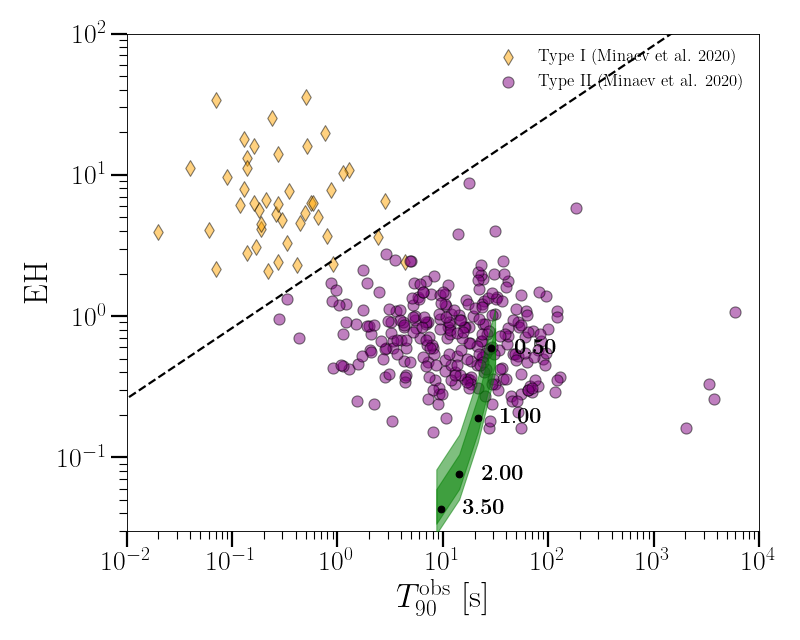}
     \caption{GRB 241113B track in the EHD classification for the redshift range z $\in [0.5 - 3.5]$ from \citet{Minaev2020}. This plot is a reproduction of the figure published by  \citet{Minaev2020}.}
     \label{fig:EHD_classification}
\end{figure}

\FloatBarrier
\newpage
\section{Multiwavelength Follow-up Observations of GRB 241113B}
\begin{table}[h]
\begin{tabular}{llllll}
\multicolumn{6}{c}{\large \textbf{X-ray}} \\ \hline \hline
\\
$T_{\rm mid} - T_{\rm GRB}$$^{(1)}$ & Exp. time & Flux$^{(2)}$  & Flux density$^{(3)}$ & Band & Telescope/Instrument \\
 (s) & (s)               & ($\mathrm{10^{-13}~erg~cm^{-2}~s^{-1}}$) &   (mJy) &   &   \\ \hline
 15604 & 1006 &  $9.7\pm 0.4$ & $6.5\pm0.2\times 10^{-5}$ & 0.3--10 keV & \textit{EP}/FXT\\ 
 16610 & 1006 &  $12.1\pm 0.5$ & $8.1\pm0.3\times 10^{-5}$ & 0.3--10 keV & \textit{EP}/FXT\\ 
 17616 & 1006 &  $8.5\pm 0.4$ & $5.7\pm0.2\times 10^{-5}$ & 0.3--10 keV & \textit{EP}/FXT\\ 
\hline
\\
\multicolumn{6}{c}{\large \textbf{Optical and NIR}} \\ \hline \hline
\\
$T_{\rm mid} - T_{\rm GRB}$$^{(1)}$ & Exp. time & Magnitude$^{(4)}$ & Flux density$^{(3)}$ & Band & Telescope/Instrument \\
(s) & (s)   & (mag) &   (mJy) &   &   \\ \hline
 13874 &  1530 & $22.72\pm0.12$ & $4.75\pm0.49 \times10^{-3}$ & $B_{VT}$ & \textit{SVOM}/VT\\ 
 19712 &  1530 & $23.21\pm0.19$ & $3.02\pm0.48 \times10^{-3}$ & $B_{VT}$ & \textit{SVOM}/VT \\
 25535 &  1500 & 23.74 $^{\dagger}$ &  $1.86 \times10^{-3}$ & $B_{VT}$ & \textit{SVOM}/VT \\
 31387 &  1530 & 23.80 $^{\dagger}$&  $1.75 \times10^{-3}$ & $B_{VT}$ & \textit{SVOM}/VT \\
288292 &  1920 & 24.09 $^{\dagger}$&  $1.34 \times10^{-3}$ & $B_{VT}$ & \textit{SVOM}/VT \\
 \hline
 18776 &  1800 & 22.05 $^{\dagger}$& $7.26 \times10^{-3}$ & $v$ & MEPHISTO \\
\hline
  245  &  10 & $16.67\pm0.04$ $^{\ddagger}$ &  $0.75\pm0.02$ & {\it Clear} & KAIT\\
  548  &  10 & $17.19\pm0.04$ $^{\ddagger}$&  $0.46\pm0.01$ & {\it Clear} & KAIT\\ 
  850  &  10 & $17.75\pm0.07$ $^{\ddagger}$&  $0.28\pm0.02$ & {\it Clear} & KAIT\\ 
 1235  &  20 & $18.34\pm0.08$ $^{\ddagger}$&  $0.16\pm0.01$ & {\it Clear} & KAIT\\ 
 1722  &  20 & $18.83\pm0.11$ $^{\ddagger}$&  $0.10\pm0.01$ & {\it Clear} & KAIT\\ 
 2202  &  20 & $19.23\pm0.14$ $^{\ddagger}$&  $7.08\pm0.84 \times10^{-2}$ & {\it Clear} & KAIT\\ 
 2771  &  30 & $19.64\pm0.23$ $^{\ddagger}$&  $4.89\pm0.93 \times10^{-2}$ & {\it Clear} & KAIT\\ 
 3436  &  30 & $19.92\pm0.24$ $^{\ddagger}$&  $3.76\pm0.75 \times10^{-2}$ & {\it Clear} & KAIT\\ 
 4292  &  30 & $20.14\pm0.19$ $^{\ddagger}$&  $3.05\pm0.49 \times10^{-2}$ & {\it Clear} & KAIT\\ 
 4931  &  30 & $20.44\pm0.31$ $^{\ddagger}$&  $2.33\pm0.58 \times10^{-2}$ & {\it Clear} & KAIT\\ 
 5570  &  30 & $20.46\pm0.33$ $^{\ddagger}$&  $2.28\pm0.59 \times10^{-2}$ & {\it Clear} & KAIT\\ 
 6219  &  30 & $20.55\pm0.39$ $^{\ddagger}$&  $2.11\pm0.64 \times10^{-2}$ & {\it Clear} & KAIT\\ 
 7132  &  100 & $20.77\pm0.35$ $^{\ddagger}$& $1.71\pm0.47 \times10^{-2}$ & {\it Clear} & KAIT\\ 
 13874 & 1470 & $22.31\pm0.10$ & $6.06\pm0.53 \times10^{-3}$ & $R_{VT}$ & \textit{SVOM}/VT \\
 18776 & 1800 & 22.55 $^{\dagger}$& $4.20 \times10^{-3}$ & $r$ & MEPHISTO \\
 19697 & 1440 & $23.06\pm0.19$ & $3.03\pm0.49 \times10^{-3}$ & $R_{VT}$ & \textit{SVOM}/VT \\ 
 25535 & 1440 & $23.37\pm0.25$ & $2.28\pm0.46 \times10^{-3}$ & $R_{VT}$ & \textit{SVOM}/VT \\ 
 31387 & 1470 & 23.42 $^{\dagger}$& $2.18 \times10^{-3}$ & $R_{VT}$ & \textit{SVOM}/VT \\ 
288292 & 2040 & 23.85 $^{\dagger}$& $1.47 \times10^{-3}$ & $R_{VT}$ & \textit{SVOM}/VT \\ 
1333852 & 2040 & 22.84 $^{\dagger}$& $2.65 \times10^{-3}$ & $GAIA_{G}$ & RAPAS \\ 
\hline
 18776 & 1800 & 21.38 $^{\dagger}$&  $1.11 \times10^{-2}$ & $z$ & MEPHISTO \\
\hline
 4860 & 3600 & 19.10 $^{\dagger}$&  $3.24 \times10^{-2}$ & $J$ & WINTER \\
\end{tabular}%
\caption{Log of the X-ray to NIR follow-up observations of GRB 241113B.}
\tablefoot{\\$^{(1)}$$T_{\rm GRB} = \mathrm{2024\text{-}11\text{-}13T11{:}23{:}07.60}$ UTC.\\
               $^{(2)}$Unabsorbed flux.\\
               $^{(3)}$The optical flux densities are corrected for the Galactic visual extinction, $A_V$, along the line of sight before being computed. The X-ray flux densities are computed from the unabsorbed spectrum at the specific energy $E_\gamma=1$ keV.\\
               $^{(4)}$The magnitudes are given in the AB or Vega (specified by $\ddagger$) photometric systems and are not corrected for the Galactic visual extinction, $A_V$, along the line of sight.\\
               $^{\dagger}$Upper limit on the flux.}
\label{tab:followup_obs}
\end{table}

\FloatBarrier
\newpage
\section{GRB 241113B Afterglow Modeling}

\subsection{Phenomelogical Fit}
\label{appendix:pheno_fit}

Results of the GRB 241113B afterglow fit with a simple smoothly broken power model; see Equation \ref{equ:aglow_pheno_model}.

\begin{table}[h]
\centering
\begin{tabular}{cccccc}
\hline
$t_b$ & $\alpha_1$ & $\alpha_2$ & $\beta$ & s & $\chi^{2}$ (dof)\\
(s) &            &            &         &    &              \\
\hline
\hline
\\
263$^{+68}_{-91}$ & 1.52$^{+1.90}_{-1.11}$ & -1.40$^{+0.03}_{-0.04}$ & 0.66$^{+0.04}_{-0.03}$ & 1.20$^{+1.02}_{-0.61}$ &   6.1 (15)\\
\\
\hline
\end{tabular}%
\caption{Best-fit model parameters. The errors are given at the 68$\%$ confidence level.}

\label{tab:pheno_fit_result}
\end{table}

\begin{figure}[h!]
    \centering
    \includegraphics[width=0.5\textwidth]{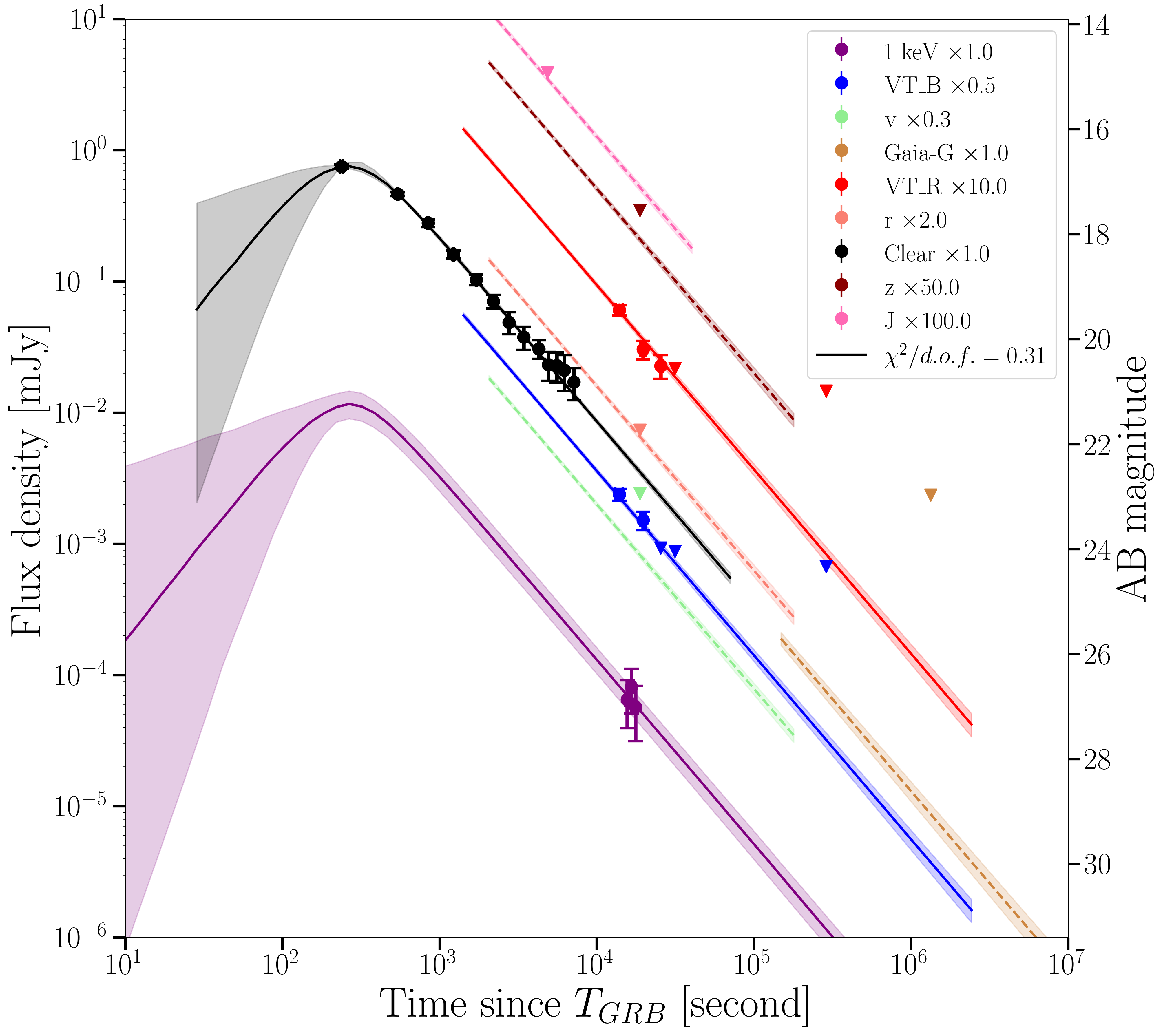}
    \includegraphics[width=0.4\textwidth]{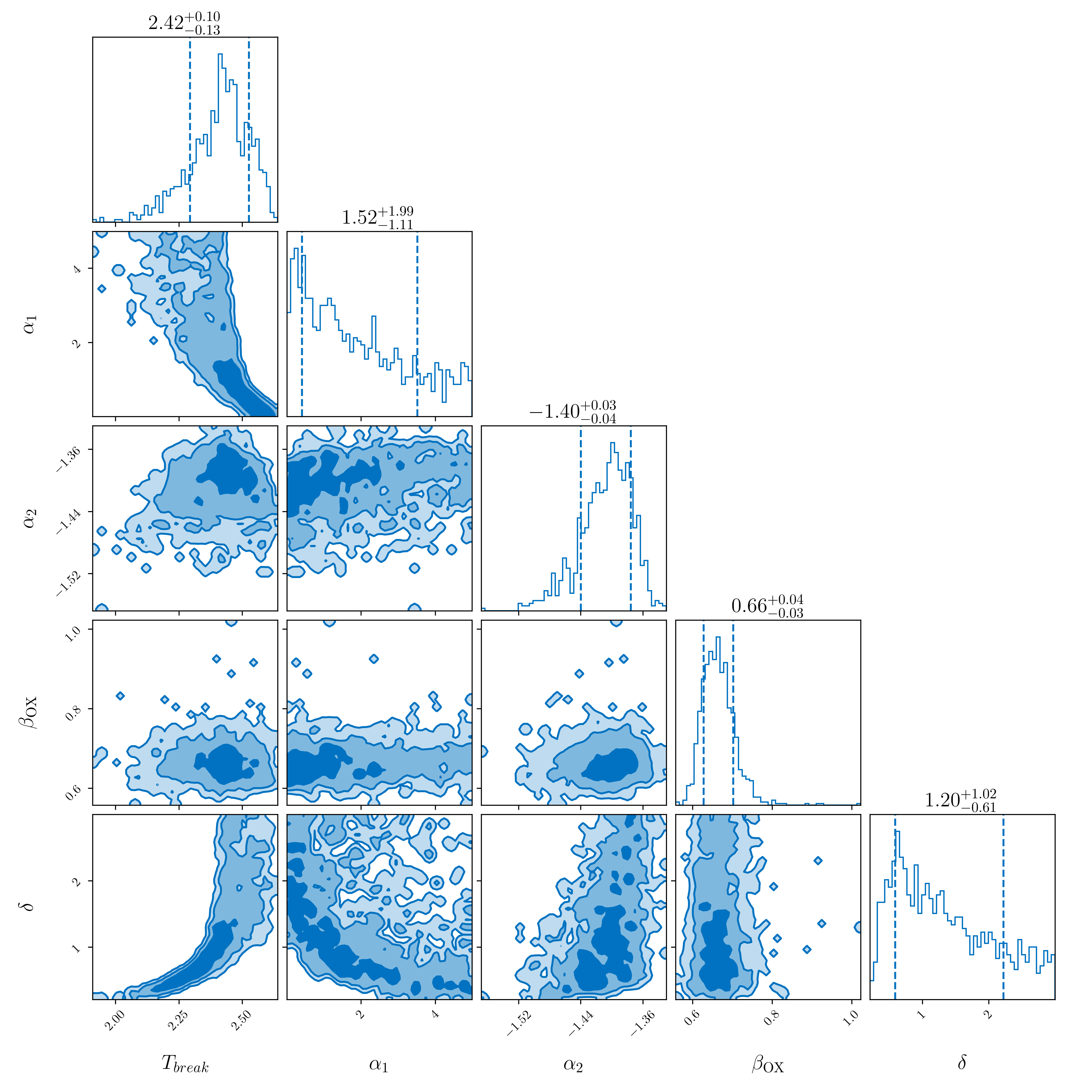}
    \caption{GRB 241113B afterglow fit with a simple phenomenological model and the posterior distribution of the model's parameters. Note that $T_{break}$ is expressed in seconds in $log_{10}$ scale.}
    \label{fig:aglow_pheno_model}
\end{figure}

\newpage
\subsection{\textsc{VegasAfterglow} Model Fit}
\label{appendix:vegasafterglow_fit}

\subsubsection{ISM}

\FloatBarrier
\begin{figure}[h!]
    \centering
    \includegraphics[width=0.45\columnwidth]{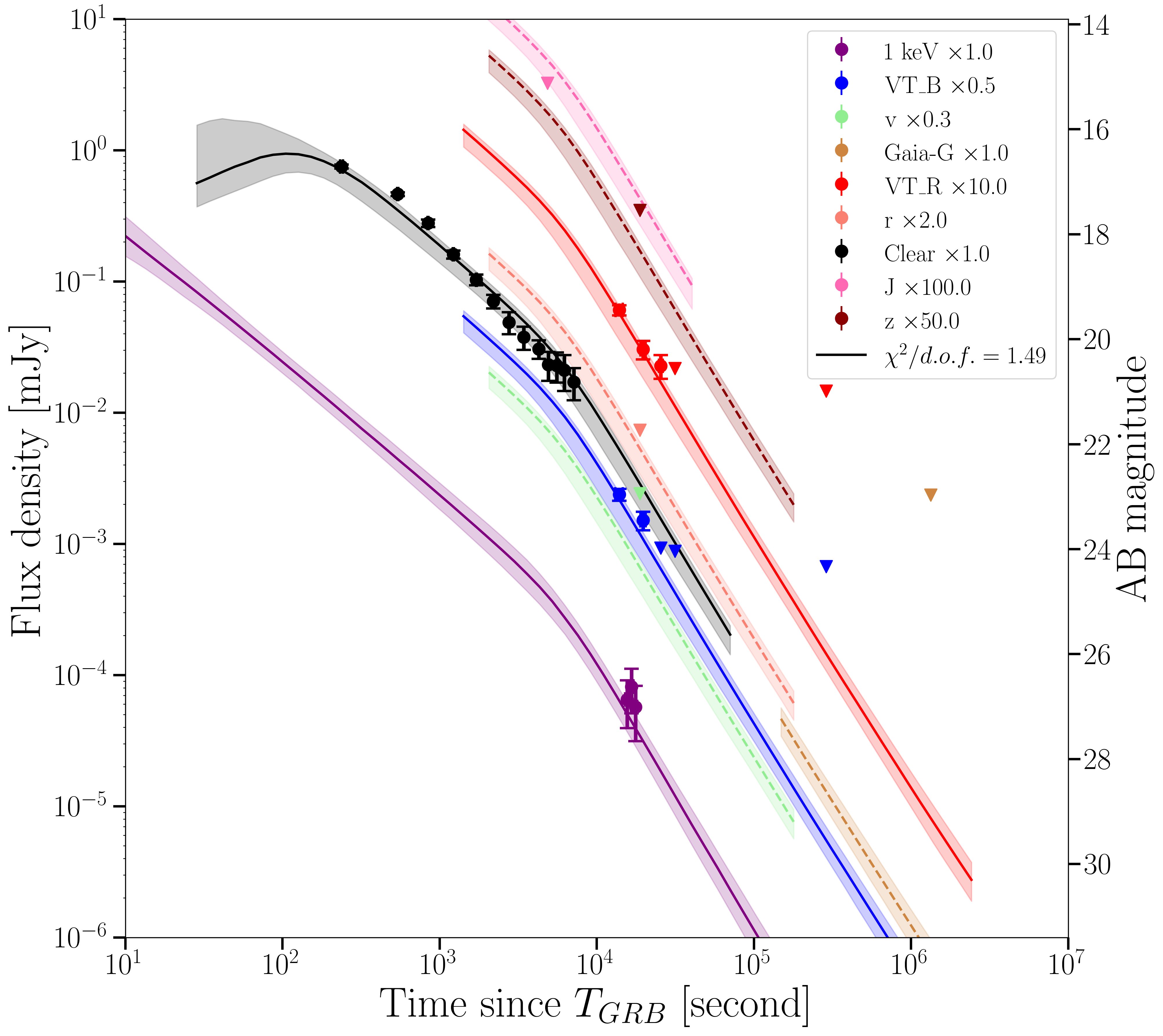}
    \includegraphics[width=0.5\textwidth]{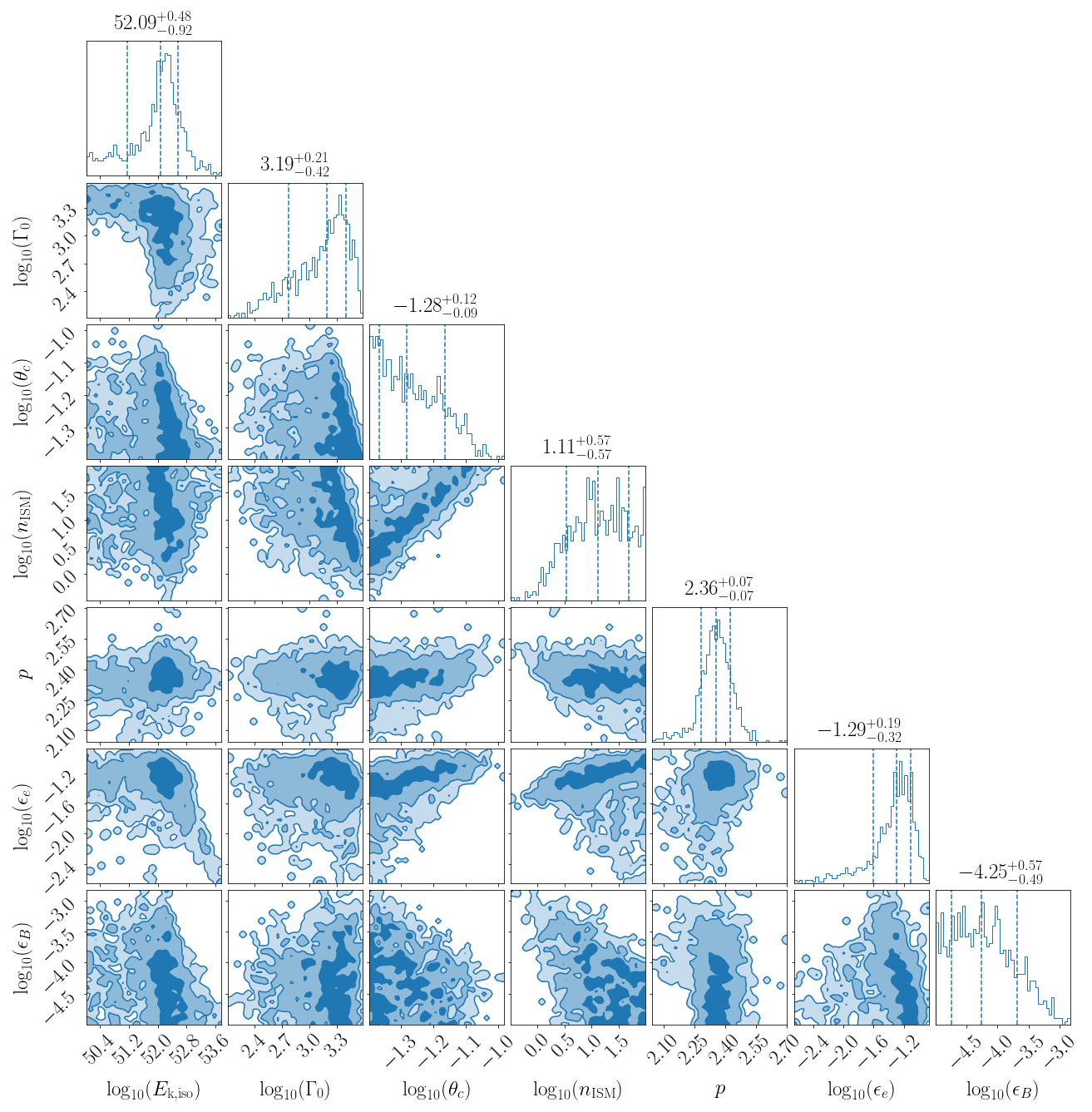}
    \caption{GRB 241113B afterglow fit with the \textsc{VegasAfterglow} code using a simple top-hat jet propagating into a constant-density ISM and at $z = 0.5$.}
    \label{fig:aglow_wind_model_z05}
\end{figure}

\FloatBarrier
\begin{figure}[h!]
    \centering
    \includegraphics[width=0.45\columnwidth]{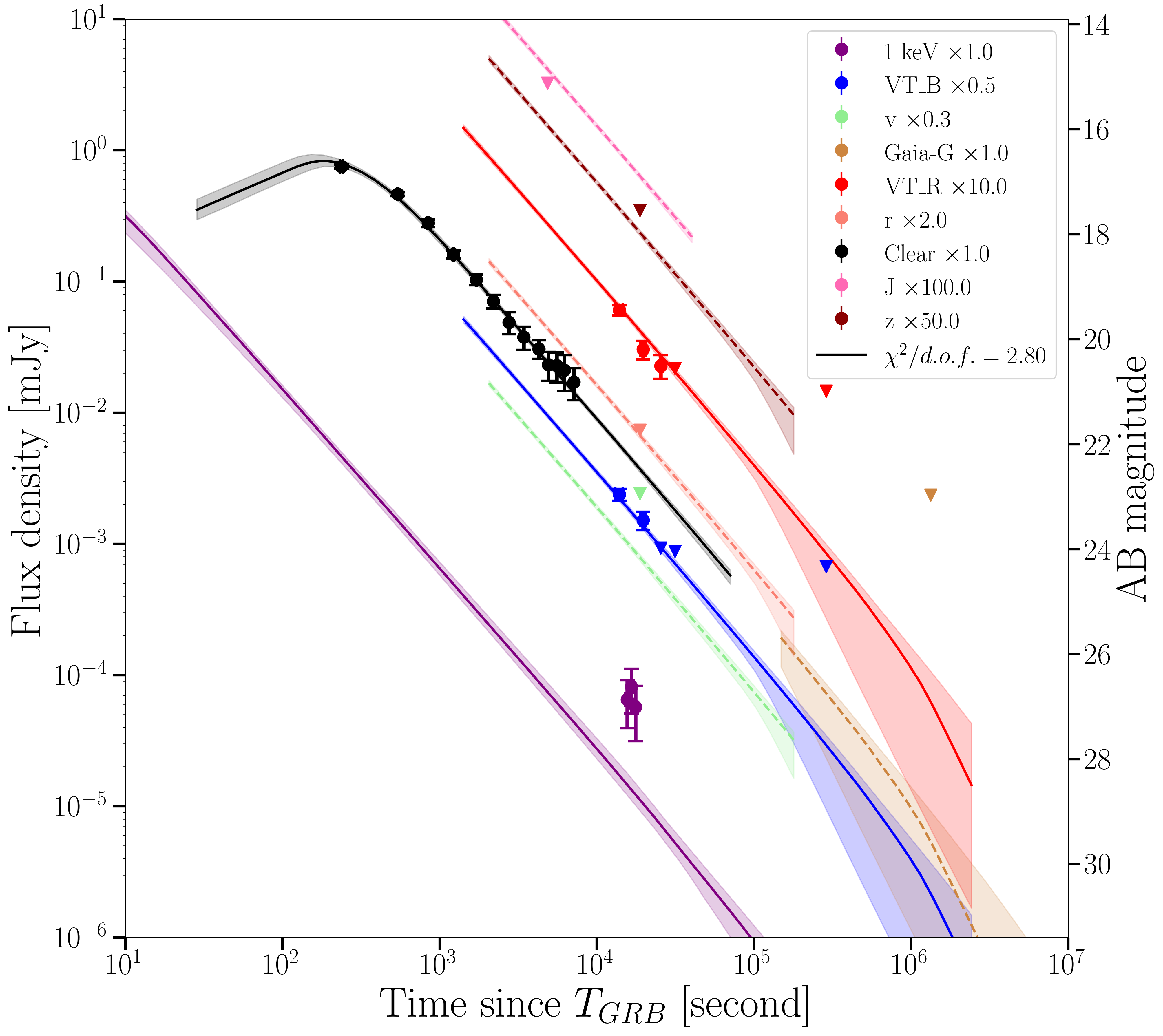}
    \includegraphics[width=0.5\textwidth]{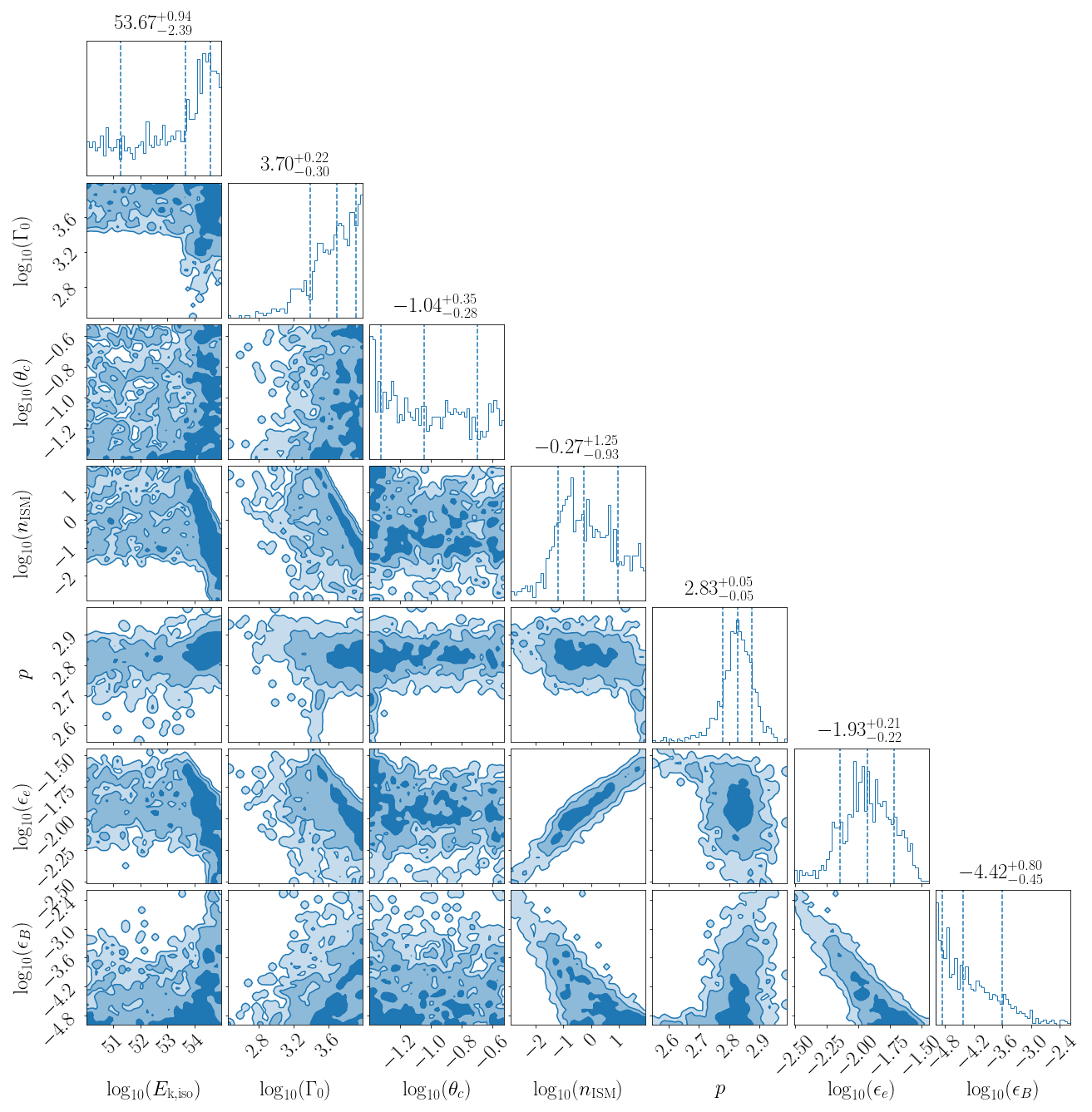}
    \caption{GRB 241113B afterglow fit with the \textsc{VegasAfterglow} code using a simple top-hat jet propagating into a constant-density ISM at $z = 3.5$.}
    \label{fig:aglow_wind_model_z35}
\end{figure}

\FloatBarrier
\subsubsection{Wind-like Medium}
\begin{figure}[h!]
    \centering
    \includegraphics[width=0.5\textwidth]{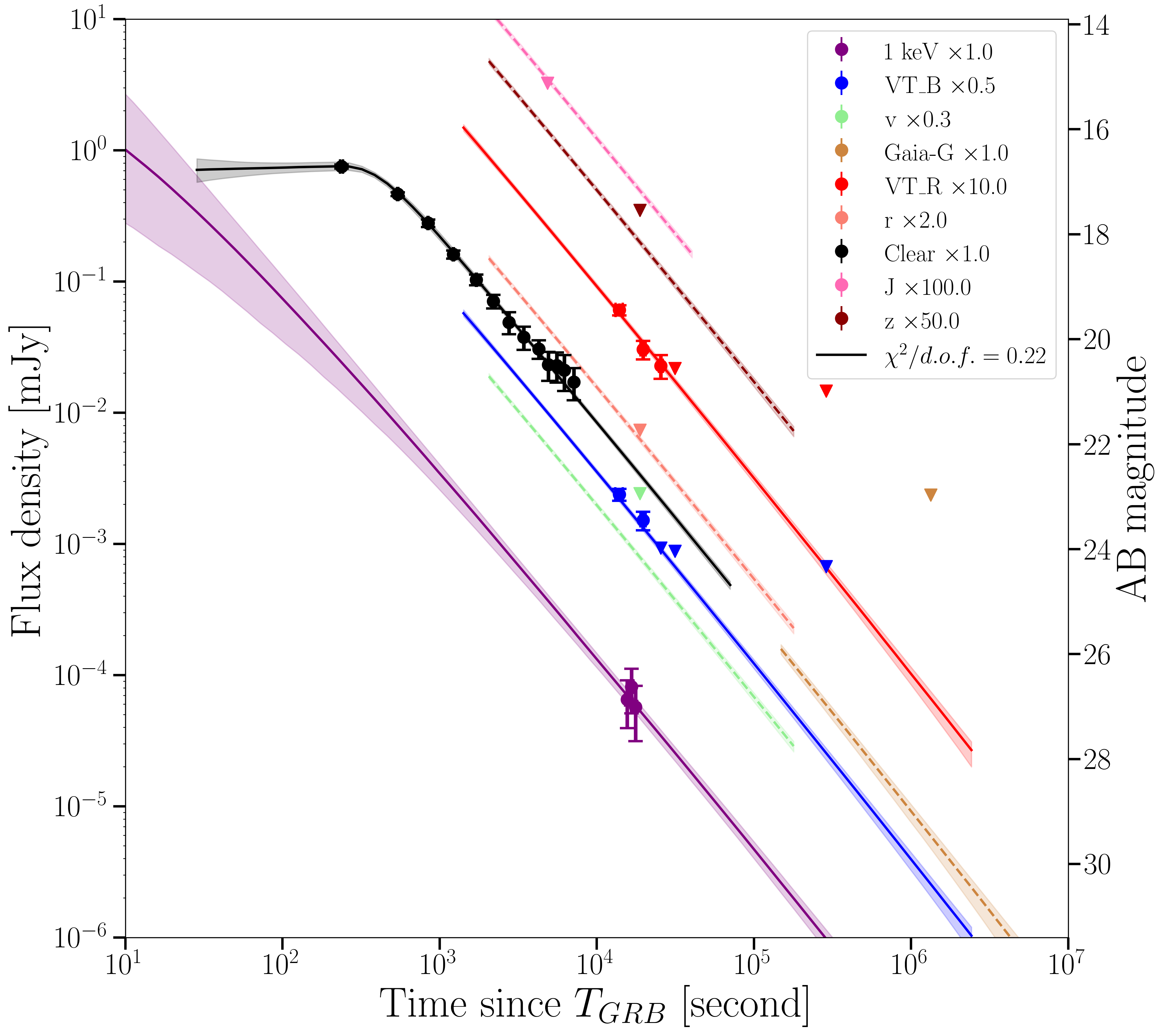}
    \includegraphics[width=0.4\textwidth]{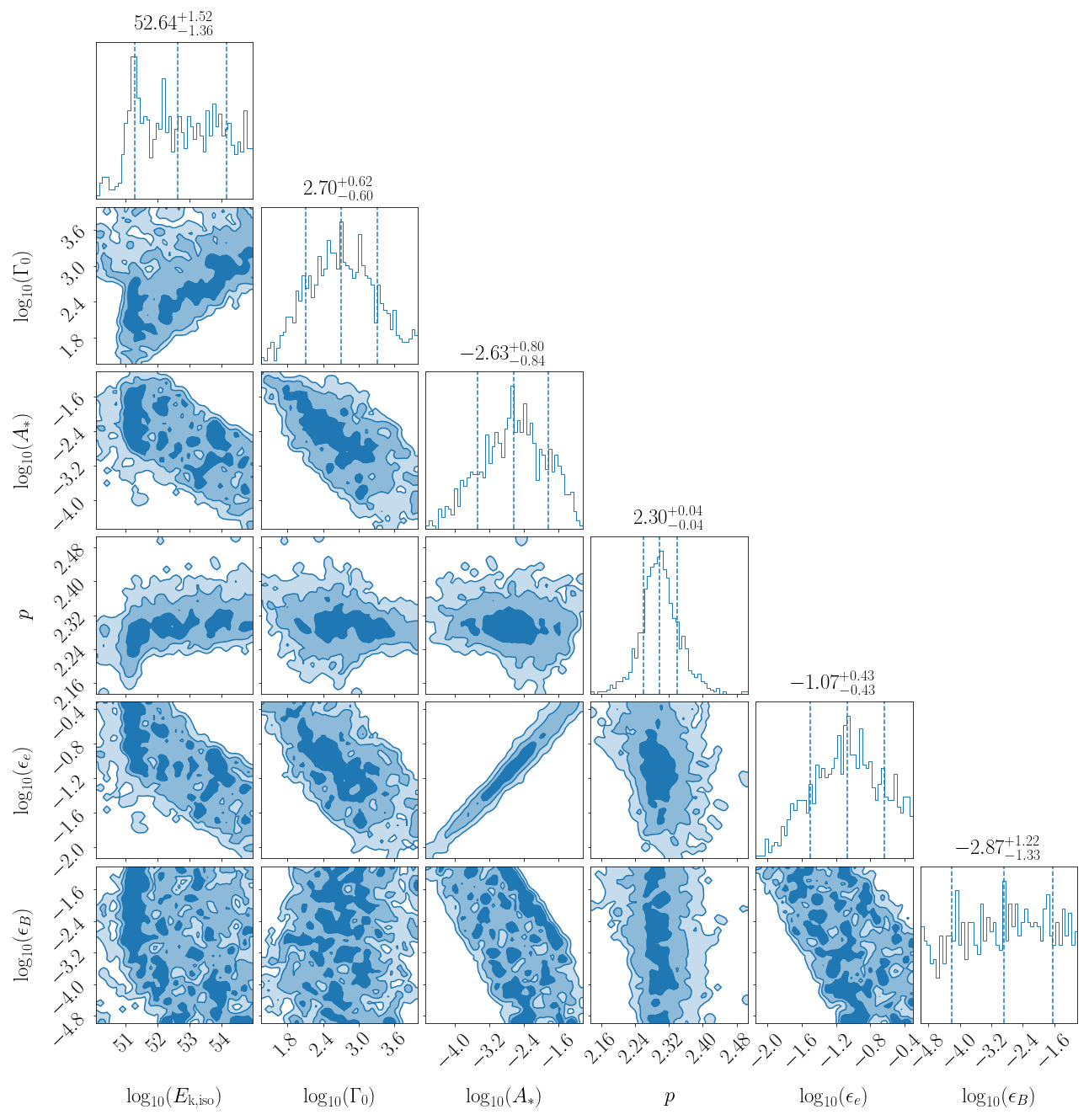}
    \caption{GRB 241113B afterglow fit with the \textsc{VegasAfterglow} code using a simple top-hat jet propagating into a wind-like medium at $z = 0.5$.}
    \label{fig:aglow_wind_model_z05}
\end{figure}

\FloatBarrier
\begin{figure}[h!]
    \centering
    \includegraphics[width=0.5\textwidth]{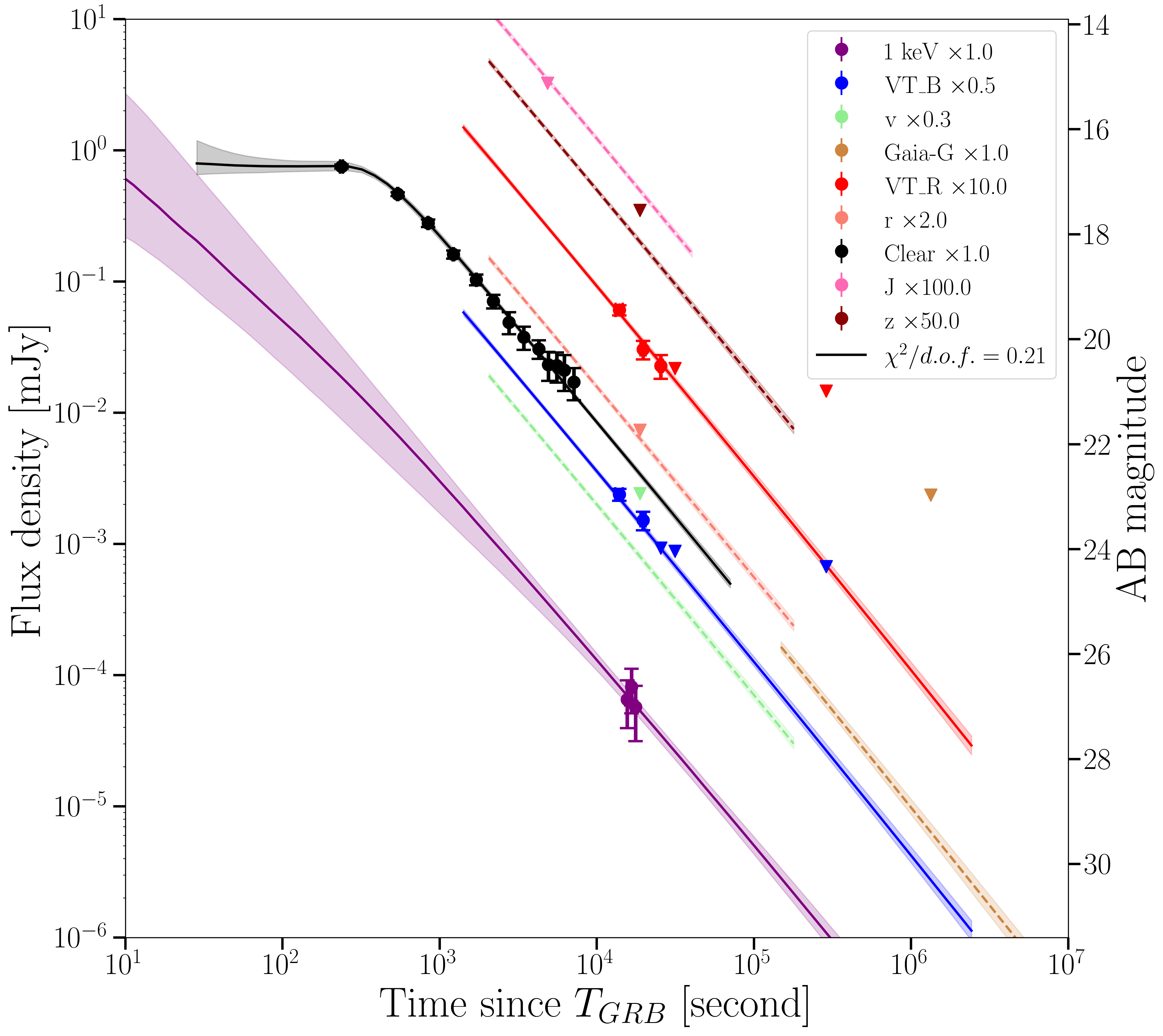}
    \includegraphics[width=0.4\textwidth]{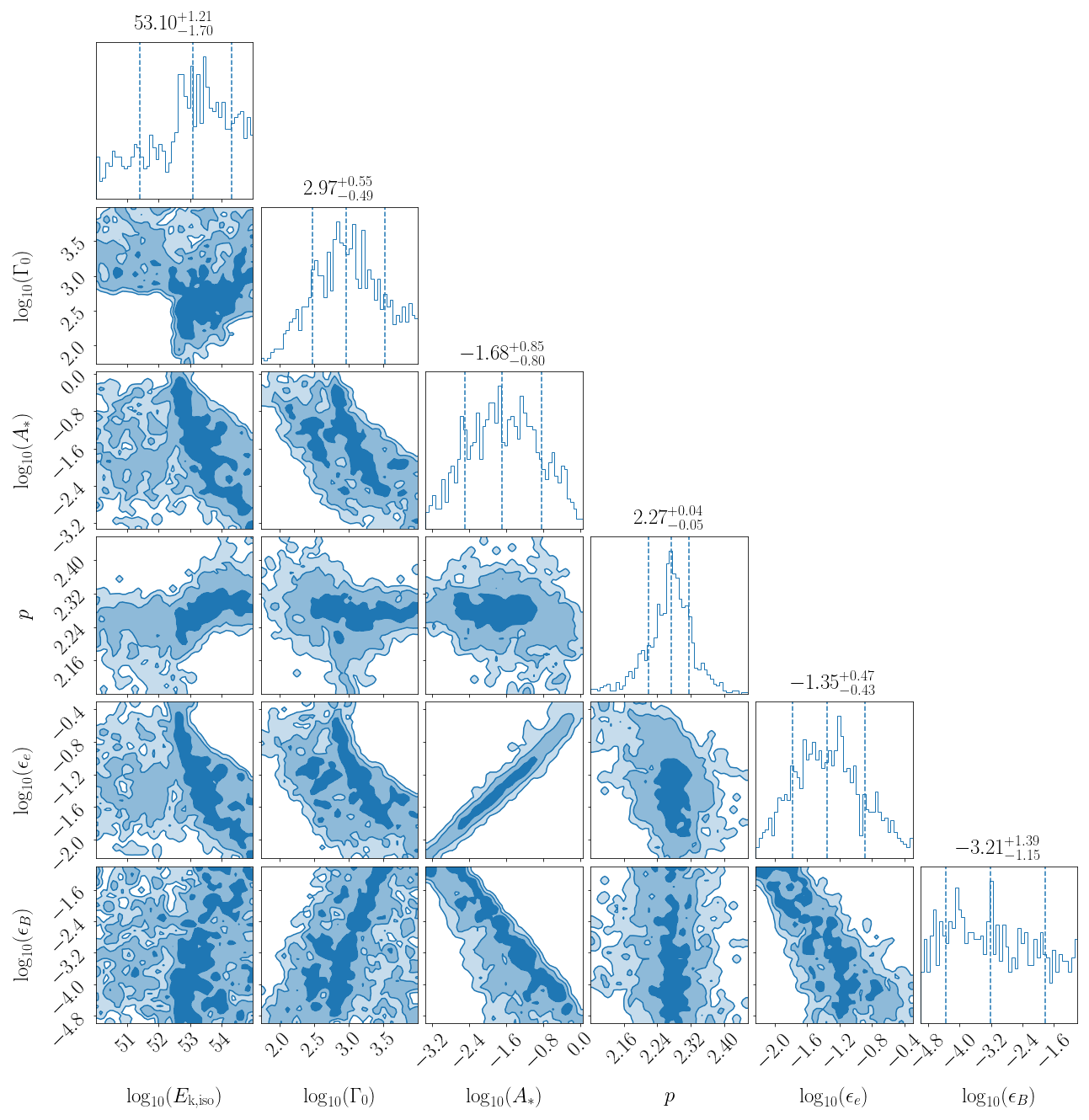}
    \caption{GRB 241113B afterglow fit with the \textsc{VegasAfterglow} code using a simple top-hat jet propagating into a wind-like medium at $z = 3.5$.}
    \label{fig:aglow_wind_model_z35}
\end{figure}

\end{appendix}